\RequirePackage{fix-cm}
\documentclass[twocolumn,epjc3]{svjour3}  
\smartqed  
\usepackage[T1]{fontenc}
\usepackage[utf8]{inputenc}
\usepackage{graphicx}
\usepackage{amsmath}
\usepackage{amssymb}
\usepackage{hyperref}
\usepackage{cite}
\usepackage{xcolor}
\usepackage[UKenglish]{babel}
\usepackage{booktabs}
\usepackage{subcaption} 
\usepackage{multirow}
\usepackage{orcidlink}
\usepackage{siunitx}

\journalname{Eur. Phys. J. C}
\begin{document}

\title{A compact, interrogable Tree Tensor Network Jet Tagger on Lund declustering trees
}


\author{Fabrizio Napolitano\thanksref{e1,addr1,addr2}\orcidlink{0000-0002-8686-5923}
        \and
        Luca Della Penna\thanksref{e2,addr1,addr2}\orcidlink{0009-0007-8978-1067}
        \and
        Tommaso Tedeschi\thanksref{e3,addr2}\orcidlink{0000-0002-7125-2905}
        \and 
        Livio Fanò\thanksref{addr1,addr2}\orcidlink{0000-0002-9007-629X}
}

\thankstext{e1}{e-mail: fabrizio.napolitano@unipg.it}
\thankstext{e2}{e-mail: luca.dellapenna@pg.infn.it}
\thankstext{e3}{e-mail: tommaso.tedeschi@pg.infn.it}

\institute{Università degli Studi di Perugia, Dipartimento di Fisica e Geologia, Via A. Pascoli, 06123, Perugia (PG), Italy \label{addr1}
\and
INFN Sezione di Perugia, Via A. Pascoli, 06123 Perugia (PG), Italy \label{addr2}
}

\date{Received: date / Accepted: date}

\maketitle

\begin{abstract}
Jet taggers built on the Lund declustering tree encode QCD radiation structure in the inputs and process it with graph networks. We ask whether the tree itself can also constrain the computation, and what becomes accessible when it does. For this, we construct a novel quantum-inspired Tree Tensor Network (TTN), where a single tensor core is shared among the internal nodes of the Cambridge/Aachen declustering tree, employing Lund coordinates, and with no particle-identification or detector-level information. The construction is based on a strong prior on the radiation history rather than on the constituents, complementary to compact equivariant taggers. Because the same learned map acts identically at every node, the trained model can be evaluated on truncated trees and its latent space compared across recursions, and used to localise where on the Lund jet plane the Pythia and Herwig jet representations diverge, showing the regions of generator-dependent concentration. The $\chi=8$ model has 4014 trainable parameters against approximately 391,000 for LundNet. We evaluate boosted top tagging and quark/gluon discrimination at particle level on Pythia and Herwig samples, in both native (Pythia$\to$Pythia) and transfer (Pythia$\to$Herwig) configurations. Despite differing in capacity by two orders of magnitude, the TTN reaches native performance close to LundNet  in quark/gluon discrimination, slightly underperforming in top tagging, with a similar Pythia-to-Herwig transfer behavior in both tasks. Finally, we show the behavior of the network as a function of the bond dimension, and as a function of the perturbative scale. 
\end{abstract}

\section{Introduction}
\label{intro}
Jet substructure provides a central handle at the Large Hadron Collider (LHC) for identifying hadronically-decaying boosted heavy particles which show multi-prong structures, and separating quarks from gluons and different flavours based on soft radiation patterns associated with the initiating colour charge~\cite{Larkoski:2017jix}. 
In recent years, machine learning (ML) taggers have become ubiquitous due to the advancement and capacity of the architectures, now occupying a central role in the experimental physics programme with jets~\cite{Kasieczka:2019dbj}. They can roughly be divided in two families.
The first family of taggers imposes exact kinematic symmetries of the four-vectors, through Lorentz and permutation equivariance~\cite{Bogatskiy:2020tje,Spinner:2024hjm,Gong:2022lye,bogatskiy2022pelican}, introducing a strong physics bias allowing competitive performance at reduced model capacity. 
The second family concerns taggers which employ low-level observables such as jets' constituents, processing them with sets, graphs, images or tokens with highly sophisticated architectures and substantial number of trainable parameters~\cite{Komiske:2018cqr,Qu:2019gqs,Qu:2022mxj}. They have the strongest raw discrimination power, at the cost of an enhanced modelling dependence and limited interpretability.
In the recent literature, increasing attention has been put on robustness of taggers with regards to both parton shower and hadronization modelling, but also associated experimental systematic uncertainties. In particular, a model trained on one parton-shower and hadronization model may lose a significant fraction of its discrimination power when evaluated on another; this loss is generally larger for more expressive models~\cite{Gambhir:2025pareto}. Systematic studies have shown this effect in quark/gluon tagging~\cite{Butter:2022xyj}, while adversarial decorrelation can underestimate the true uncertainty~\cite{Ghosh:2021roe}. Recent studies~\cite{SciPostPhys.17.6.166} have shown that low-level inputs translate directly into a higher modelling uncertainty, as high as 40\%.
The Lund jet plane offers a theoretically clean representation of the jet~\cite{Dreyer:2018nbf}, giving direct access to a small set of IRC-safe kinematic observables, where different regions of the plane are populated by different QCD behaviors. The Lund plane has been used for tagging since its introduction~\cite{Dreyer:2018nbf,Patel:2026xai,dreyer2022quarks}: LundNet~\cite{Dreyer:2020brq} has been shown to achieve competitive performance exploiting the declustering structure via a large edge-convolutional graph network. 
A further advantage of this representation is that the primary Lund plane density is itself a measured observable, unfolded to particle level by ATLAS, CMS and LHCb~\cite{ATLAS:2020bbn,CMS:2023lpp,LHCb:2025mcq}, with its uncertainty and covariance information. The remaining inputs of
Lund-tree taggers (secondary branches, $\ln m$, $\psi$) are not directly
measured, but are constrained indirectly by measured substructure
observables. The architecture of these taggers, however, uses the declustering tree as a graph for message passing, but does not constrain the computation to follow the underlying branching structure.

Quantum-inspired tensor networks provide a natural class of models for tree representation. Originating in quantum many-body physics as compact parametrization of states with limited entanglement~\cite{White:1992zz,Cirac:2020obd,Vidal:2008zz,Shi:2006zz}, they have been adopted as parameter-efficient classifiers whose capacity is controlled by a single bond dimension ~\cite{Stoudenmire:2016dbs}, making them attractive in HEP for two reasons: (i) small parameter count and structured contraction, important for low latency implementations; and (ii) the bond dimension is an interpretable parameter, which controls the capacity of information the model transmits between subtrees.
Tensor networks have been applied to HEP data for b-tagging at LHCb~\cite{felser2021quantum} and for event classification~\cite{araz2021quantum,araz2022classical}.
Tree Tensor Network classifiers have recently been developed for low-latency jet tagging with FPGA, demonstrating competitive performances and inference time well below that of graph or transformer architectures~\cite{Coppi:2026ttn}. In that work the jet's constituents enter the network through the leaves, and the network topology is a design choice made for computational efficiency, not tied to the physical structure of the jet.

We take a  different approach, constructing the network itself along the jet declustering tree. At each node, the associated Lund plane features are computed with a single low-bond-dimension tensor core which is shared across every internal node. The model therefore learns a splitting map across the scales of the QCD radiation history, enforced by the architecture. 
In this way we employ the Lund plane observables, while processing them in a representation which is physically inspired by the structure of the jet. 
Like other Lund-tree taggers, and unlike unconstrained constituent-level taggers such as ParticleNet, the construction has a direct handle on IRC safety with a $k_t$ cut on the tree. 
This construction also builds on our previous quantum tree-topology networks, based on Lund Plane to Bloch encoding~\cite{Napolitano:2026lp2b}, but restructuring the architecture faithfully to (classical) tensor networks.

Processing a jet along the clustering history recursively has already been pioneered in the work of Louppe et al.~\cite{Louppe:2017ipp}, where leaf embeddings of the constituents four-momenta are combined in pairs by a shared weight matrix. A related construction~\cite{Andreassen:2018apy} learns a shared per-splitting model on the tree, providing interpretable factorization of the likelihood.
Our architecture retains the shared model of the branchings, but differs in: (i) the tensor approach is multilinear rather than affine, and the correlations between siblings are represented in the functional form itself; (ii) the node features are IRC-safe Lund coordinates, rather than constituents four-momenta; (iii) the capacity of the model is set by a single bond dimension $\chi$, with a direct interpretation about the amount of information transmitted between subtrees.

The remainder of the paper is organized as follows: in 
Section~\ref{sec:ttnmethod} we introduce the methodology and the architecture of the TTN on the Lund declustering tree; in Section~\ref{sec:data} we introduce the dataset used in this work; in Section~\ref{sec:training} we show the training setup and benchmarks; in Section~\ref{sec:results} we present the results and in Section~\ref{sec:conlusions} the conclusions.

\section{Tree Tensor Network method}
\label{sec:ttnmethod}

The classifier is defined on a binary declustering tree
\(\mathcal{T}=(V,E)\) obtained from the Cambridge/Aachen algorithm. Each internal node \(v\in V\)
corresponds to a resolved splitting of a parent
pseudojet into two children \(a\) and \(b\), ordered such that
\(p_{\mathrm{T},a}\geq p_{\mathrm{T},b}\). The local Lund-coordinate feature
vector is
\begin{equation}
  x_v =
  \left(
    \ln k_{t},
    \ln\frac{R_0}{\Delta},
    \ln z,
    \ln m,
    \psi
  \right)_v ,
  \label{eq:lund-features}
\end{equation}
where \(k_{t}=p_{\mathrm{T},b}\Delta\),
\(z=p_{\mathrm{T},b}/(p_{\mathrm{T},a}+p_{\mathrm{T},b})\),
\(\Delta=\sqrt{(y_b-y_a)^2+(\phi_b-\phi_a)^2}\),
\(m=m(a+b)\), and \(\psi=\operatorname{atan2}(y_b-y_a,\phi_b-\phi_a)\).
The coordinates follow the Lund-plane representation of resolved emissions~\cite{Dreyer:2018nbf}.
Splittings below the \(k_{t}\) threshold are skipped by
following the harder branch; this threshold is absent in the default implementation,
\(\ln k_{t}>-\infty\). 
Without this cut, which is the default for the results of Sec.~\ref{subsec:res_1}-\ref{subsec:res_2} and ~\ref{subsec:res_4}, the tree and hence the classifier are not IRC safe; Sec.~5.3 studies the IRC-safe regime.
The reference angular scale is \(R_0={1}\).

\begin{figure*}[htpb]
    \centering
    \includegraphics[width=0.75\textwidth]{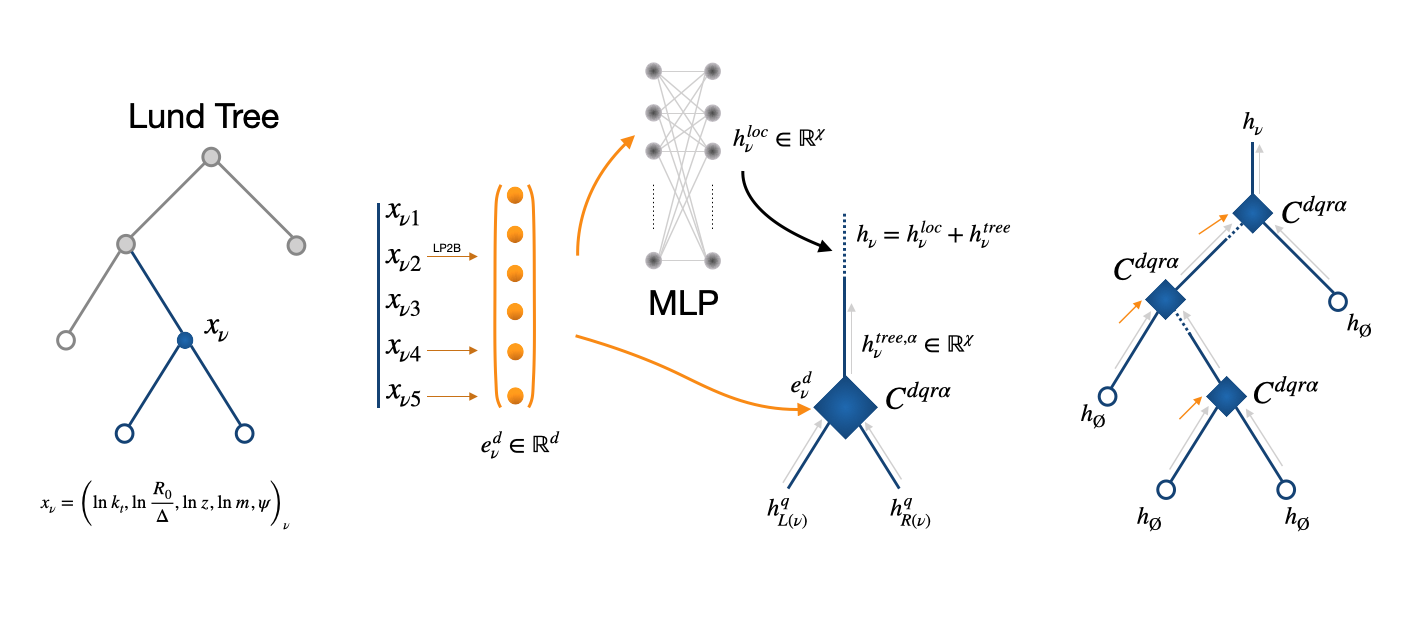}
    \caption{A schematic view of the TTN architecture. Each jet is read as a C/A declustering tree, encoded locally by a shared map, and fused by a single tensor core reused recursively at every node. The schematic shows the Lund tree (left), a binary C/A declustering tree; at each splitting node $v$ the two prongs are $p_T$-ordered into a harder child $a$ and a softer child $b$ ($p_{T,a}\!\ge\!p_{T,b}$). Every node carries Lund feature vector $x_v=(\ln k_t,\ \ln\tfrac{R_0}{\Delta},\ \ln z,\ \ln m,\ \psi)$ built from the splitting kinematics, with the transverse momentum $k_t=p_{T,b}\Delta$, momentum fraction $z$, angular separation $\Delta$, and azimuthal orientation $\psi$, using directly controllable IRC-safety through a $k_t$ cut. The feature vector is then constructed (central, orange) using the projective sphere encoding (LP2B~\cite{Napolitano:2026lp2b}) for $\ln k_t,\ \ln\tfrac{R_0}{\Delta},\ \ln z$ and the bounded $\tanh(\ln m/5)$ and $\psi/\pi$ for the remaining two features.
  $e_v$ is then used directly into the core tensor contraction and as residual via a single multilayer perceptron (central, gray) $\varphi$, with weights shared across all populated nodes.
  Node states are propagated up the tree by a recursive contraction with one shared core tensor $C^{dqr\alpha}$. At each node the tree message is $h^{{tree},\alpha}_v=\sum_{d,q,r}C^{dqr\alpha}\,e^d_v h^q_{L(v)}\,h^r_{R(v)}$, combining the encoded feature $e_v$ with the left- and right-child bonds $h_{L(v)},h_{R(v)}$ into an output bond of dimension $\chi$; missing children are replaced by a learned vacuum state $h_\varnothing$. Local map  contributes through the residual addition $h_v=h^{{tree}}_v+h^{{loc}}_v$. This building block is then repeated recursively for every node of the Lund binary tree (right). The root final bond $h_{\mathrm{root}}$ is passed to a readout MLP classifier.}
    \label{fig:ttn_schematics}
\end{figure*}

A schematic representation of the TTN architecture is shown in Figure~\ref{fig:ttn_schematics}.
\subsection{Declustering tree representation}

The TTN maps each local feature vector \(x_v\) to an encoded vector
\(e_v\in\mathbb{R}^{d}\). The default encoding uses the first three entries of Eq.~\ref{eq:lund-features} as projective sphere encoding,
\[
  e_v =\left[
  \frac{(1,s_1 x_{v,1},s_2 x_{v,2},s_3 x_{v,3})}
  {\sqrt{1+\sum_{i=1}^{3}s_i^2 x_{v,i}^2}},
  \;
  \tanh(x_{v,4}/5),
  \;
  x_{v,5}/\pi
  \right]
\]
where the three scales $s_i$ are learned. Thus the five input features
produce a six-dimensional encoded vector, $d=6$.

Each node carries a hidden bond vector \(h_v\in\mathbb{R}^{\chi}\), where
\(\chi\) is the TTN bond dimension. The default working configuration uses \(\chi={8}\), with a scan over
\(\chi\in\{{2},{4},{6},{8},{12},{16}\}\) presented in Section~\ref{subsec:res_2}.
For a node \(v\) with left and right children \(L(v)\) (harder) and \(R(v)\) (softer), the
shared-core contraction is
\[
  h_v^{\mathrm{tree},\alpha}
  =
  \sum_{d,q,r}
  C^{d q r \alpha}
  e_v^d
  h_{L(v)}^q
  h_{R(v)}^r ,
\]
where \(C\in\mathbb{R}^{d\times\chi\times\chi\times\chi}\) is shared across
all internal nodes. If a child is absent, \(h_{L(v)}\) or \(h_{R(v)}\) is
replaced by a learned normalized vacuum vector \(h_{\varnothing}\). 
Unlike the affine approach of~\cite{Louppe:2017ipp}, where the child states enter a shared weight matrix through concatenation and interact only through the subsequent nonlinearity, the shared-core contraction above is trilinear in $e_v$, \(h_{L(v)}\) and \(h_{R(v)}\); the product  $h_{L(v)}^q h_{R(v)}^r$ appears explicitly, with $\chi$ controlling the rank at which these correlations are retained.
A shared two-layer multilayer perceptron (the local map) supplies a local residual,
\[
  h_v^{\mathrm{loc}}
  =
  W_2\,\operatorname{GELU}(W_1 e_v+b_1)+b_2 ,
\]
with hidden dimension \({16}\). 
At splitting node, the complete state is:
\[
h_v
=
h_v^{\mathrm{tree}}
+
h_v^{\mathrm{loc}}.
\]

At a terminal node, $h_v=h_{\varnothing}$. The root state
$h_{\mathrm{root}}$ is passed to a two-layer readout multilayer
perceptron with hidden dimension 64, producing a single logit
$f_\theta(j)$ for jet $j$.

Two optional multiplicative gates encode perturbative QCD information
without changing the tree topology. The QCD gate computes the feature
vector
\begin{equation} 
\begin{split} 
F_v = \biggl( &\ln\frac{\alpha_s(Q_v)}{\alpha_s^{\mathrm{ref}}}, \; \tanh\frac{\ln P_{q\to qg}(z_v)}{5}, \; \\ \tanh\frac{\ln P_{g\to gg}(z_v)}{5}, 
&\;\tanh\frac{\ln P_{g\to q\bar q}(z_v)}{5} \biggr), 
\end{split} 
\end{equation}

where the splitting kernels are evaluated at the node momentum fraction
$z_v$. 
The coupling is evaluated at one loop with $n_f = \,$ 5 active flavours and
$\Lambda = 0.2$~GeV, at the scale $Q_v = \max(k_{t,v},\,1~\mathrm{GeV})$,
and clipped to $[0.05,\,1]$; all splittings with $k_{t,v} < 1$~GeV thus use
the frozen value $\alpha_s(1~\mathrm{GeV})$. The reference value
$\alpha_s^{\mathrm{ref}}$ is 0.3. The same prescription is used in the
Sudakov gate, Eq.~(4).
The QCD gate produces one positive factor per bond component,
\begin{equation}
g_{\mathrm{QCD}}^\alpha(v)
=
\exp\!\left[
\operatorname{clip}
\left(
\sum_{i=1}^{4}W_{i\alpha}F_{v,i}+b_\alpha,\,-2,\,2
\right)
\right]
\label{eq:ttn-qcd-gate}
\end{equation}
which is a multiplicative FiLM-style modulation
\cite{Perez:2018film}. Its weights and biases are initialized to zero,
so $g_{\mathrm{QCD}}^\alpha(v)=1$ at initialization.

The Sudakov gate is a scalar factor,
\begin{align}
R_v &=
\frac{\alpha_s(Q_v)}{\pi}
\left(\ell_{L(v)}^2+\ell_{R(v)}^2\right),\\
g_{\mathrm{S}}(v)
&=
\exp\!\left[
\operatorname{clip}
\left(
b_{\mathrm{S}}
-\operatorname{softplus}(a_{\mathrm{S}})R_v,\,-2,\,2
\right)
\right],
\label{eq:ttn-sudakov-gate}
\end{align}
where $\ell_c = \ln(\Delta_v/\Delta_c) \geq 0$, for $c \in \{L(v),R(v)\}$, is the angular-log interval from the parent splitting to the child splitting. The interval is set to zero for a terminal child. 

The gates act on the tree message before the local residual is added, so
that the complete state at a splitting node becomes
\begin{equation}
  h_v^{\alpha} = g_S(v)\, g_{\rm QCD}^{\alpha}(v)\, h_v^{\mathrm{tree},\alpha}
               + h_v^{\mathrm{loc},\alpha} .
\end{equation}
The local residual $h_v^{\mathrm{loc}}$ is not gated. With both gates
disabled, $g_S = g_{\rm QCD}^{\alpha} = 1$ and the state reduces to
$h_v = h_v^{\mathrm{tree}} + h_v^{\mathrm{loc}}$. Both gates are enabled in
the default configuration and are treated as ablation targets.


The training objective is binary focal loss with \(\gamma={2}\)
\cite{Lin:2017focal}. For a batch \(\mathcal{B}\), label
\(y_j\in\{0,1\}\), predicted probability \(p_j=\sigma(f_\theta(j))\), and
\(p_{t,j}=y_jp_j+(1-y_j)(1-p_j)\), the minimized loss is
\begin{equation}
  \mathcal{L}(\theta) 
  =
  \frac{1}{|\mathcal{B}|}
  \sum_{j\in\mathcal{B}}
  -(1-p_{t,j})^\gamma
  \left[
    y_j\log p_j + (1-y_j)\log(1-p_j)
  \right].
  \label{eq:focal-loss}
\end{equation}
Parameters are optimized with Adam updates \cite{Kingma:2014adam}, global gradient clipping at \({1}\), selective weight decay, and a warmup-cosine learning-rate schedule. The default batch size is \({1024}\), and the default training budget is \({50}\) epochs. The best checkpoint is selected by validation area under the ROC curve unless an operating-point rejection metric is chosen explicitly.

\section{Dataset and simulations}
\label{sec:data}

Two binary classification tasks are considered. In boosted top tagging, the signal class is hadronically decaying top jets and the background class is QCD jets. In quark/gluon discrimination, the signal class is quark-initiated jets and the background class is gluon-initiated jets. All samples used in this study are particle-level event-generator samples without detector simulation.

For boosted top tagging, the samples are taken from the Pythia8~\cite{bierlich2022comprehensive} and Herwig7~\cite{bellm2016herwig} Boosted Top and QCD Jet datasets \cite{Gambhir:2025pareto, Gambhir:2025topqcd}.
The samples in the dataset were generated with Pythia \texttt{8.331} and Herwig \texttt{7.3.0}; the jets clustered with FastJet \texttt{3.3.0}~\cite{cacciari2012fastjet} using the anti-\(k_{t}\)~\cite{cacciari2008anti} algorithm with radius \(R={0.8}\), satisfying \(p_{\mathrm{T}}\in[\SI{500}{\GeV},\SI{550}{\GeV}]\) and
\(|\eta|<{2.5}\). We use the full statistics of 2,000,000 events per generator.

For quark/gluon discrimination, the Pythia sample is the Komiske-Metodiev-Thaler Pythia \texttt{8.226} quark and gluon dataset, and the Herwig sample is the corresponding Herwig \texttt{7.1.4} quark and gluon dataset \cite{Komiske:2018cqr,Komiske:2019jim,Pathak:2019herwig}. Both samples correspond to \(pp\) collisions at \(\sqrt{s}=\SI{14}{\TeV}\), with quark and gluon jets recoiling against a \(Z\) boson decaying to neutrinos. Jets were clustered with FastJet \texttt{3.3.0} using the anti-\(k_{t}\) algorithm with \(R={0.4}\), and satisfy \(p_{\mathrm{T}}\in[\SI{500}{\GeV},\SI{550}{\GeV}]\) and \(|y|<{1.7}\). 
Also in this case we use the full statistics of the dataset, 2,000,000 events per generator. We excluded for this study processes with charm and bottom quark-jets.

The released arrays contain per-particle transverse momentum, rapidity, azimuthal angle, and particle-identification code. The TTN preprocessing uses only the kinematic entries. Zero-padded particles are removed, and the remaining constituents are reclustered with the Cambridge/Aachen algorithm to obtain the declustering tree. At each resolved splitting, the harder branch is assigned as child \(a\), the softer branch as child \(b\), and the features are the five Lund coordinates defined in Eq.~\ref{eq:lund-features}. 

No sample-level standardization is applied before the local projective-sphere encoding; scale factors in the first three Lund coordinates are trainable parameters of the classifier.

For cross-generator runs, the training and validation subsets are drawn from one generator, while the test subset is drawn from the other generator with the same event count as the corresponding held-out test subset. The default top-tagging split fractions are \({0.70}\), \({0.15}\), and \({0.15}\), giving \({1,400,000}\), \({300,000}\), and \({300,000}\) jets for training,
validation, and testing when all files are used. The default quark/gluon split fractions are \({0.80}\), \({0.10}\), and \({0.10}\), giving \({1,600,000}\), \({200,000}\), and \({200,000}\) jets when all files are used. 


\section{Training setup and benchmarks}
\label{sec:training}

The default model uses the five-feature Lund input, the projective-sphere local encoding, bond dimension \(\chi={8}\), a shared two-layer local map, a shared TTN core, QCD and Sudakov gates, and a root-only readout. The local encoded dimension is \(d={6}\). The local map has hidden dimension
\({16}\), and the readout multilayer perceptron has hidden dimension \({64}\). With these settings the model contains \({4014}\) trainable parameters. The parameter count is decomposed as \({11}\) vacuum and encoding-scale parameters, \({248}\) local map parameters, \({3072}\) core-tensor parameters, \({42}\) physics gates and \({641}\) readout parameters.
The shared-core tensor is initialized from a normal distribution with standard deviation \(1/\sqrt{d\chi^2}\). The local map and readout weights are initialized with fan-in scaling, and all corresponding biases are initialized to zero. The projective-sphere coordinate scales are initialized to \({0.7}\). The learned vacuum state is initialized along the first bond coordinate and normalized before use. 

Training uses the focal loss with \(\gamma={2}\).
The optimizer is implemented with Optax \cite{deepmind:2020jax} and consists of global gradient clipping at norm \({1}\), Adam moments, selective weight decay, and a scheduled learning rate. The weight decay coefficient is \({1e-5}\). Weight decay is applied to the local map weights, the TTN core tensor, and the readout weights, but not to the local encoding parameters, biases, the vacuum state, or the gate parameters. For the default batch size \({1024}\), the learning rate follows a warmup-cosine schedule with initial value \({1e-3}\), peak value \({1e-2}\), warmup length \({10}\) epochs, and final value \({1e-3}\). The default training budget is \({50}\) epochs.

The LundNet baseline reported here uses six convolution blocks with channel widths \(({32},{32})\), \(({32},{32})\), \(({64},{64})\), \(({64},{64})\),
\(({128},{128})\), and \(({128},{128})\), followed by a fully connected layer of width \({256}\) with dropout probability \({0.1}\). No minimum \(\ln k_{t}\) or \(\ln\Delta\) cut is applied. In the original paper, it is referred to as LundNet5 when processing the inputs  $\ln k_t, \ln\frac{R_0}{\Delta},\ln z,\ln m, \psi$.

The ParticleNet baseline reported here uses three neighborhood-convolution stages with \({16}\) neighbors and channel widths
\(({64},{64},{64})\),
\(({128},{128},{128})\), and
\(({256},{256},{256})\), followed by a fully connected layer of width \({256}\) with dropout probability \({0.1}\). No minimum
\(\ln k_{t}\) or \(\ln\Delta\) cut is applied.

LundNet and ParticleNet are trained with identical settings as in~\cite{Dreyer:2020brq}.

Both baselines use the same kinematic inputs as the TTN, with no
particle-identification information. The ParticleNet configuration has
365,578 trainable parameters, and LundNet5 approximately 391,000.

\section{Results}
\label{sec:results}

We show in this Section the results for quark/gluon and top/QCD
discrimination. All models are trained on Pythia, the nominal generator in
LHC practice, and evaluated on Pythia (native) and Herwig (transfer).
Following~\cite{Gambhir:2025pareto}, we quantify generator dependence by
$|\mathrm{AUC}_{P\to P}-\mathrm{AUC}_{P\to H}|$ or its relative value; this
operational figure of merit does not separate the generalisation of the
tagger from intrinsic differences in class separability between the two
samples.
In Subsection~\ref{subsec:res_1}, we show the performance of the baseline TTN model, with the LundNet and ParticleNet references. Subsection~\ref{subsec:res_2} shows the capacity scan of the bond parameter $\chi$ of the TTN; in Subsection~\ref{subsec:res_3} the scan as a function of the scale $\ln(k_t/\text{GeV})$ cut, probing the perturbative scale of the retained emissions. Finally, in Subsection~\ref{subsec:res_4} we exploit the shared core to localise changes in the latent representation along declustering recursion, and across the Lund plane. 

\subsection{Baseline Result}\label{subsec:res_1}

We show here the Receiver Operating Characteristic (ROC) curve for both tasks, for the default configuration of the TTN and the benchmarks: the directly comparable LundNet which operates on the same declustering tree, but with 391,000 parameters, and ParticleNet, which operates directly on the low-level constituents (365,578 parameters).
The AUC with its DeLong statistical uncertainty and the background rejection is shown also in Table~\ref{tab:classifier-performance}.

\begin{table*}[t]
  \centering
  \caption{Classifier performance on native (Pythia$\to$Pythia) and transfer (Pythia$\to$Herwig) samples.}
  \label{tab:classifier-performance}
  \begin{tabular}{lllccccc}
    \toprule
    Task & Evaluation & Model & AUC $\pm$ DeLong s.e. &  $1/\epsilon_B$ at $\epsilon_S=0.3$ & $1/\epsilon_B$ at $\epsilon_S=0.5$ & $1/\epsilon_B$ at $\epsilon_S=0.7$ \\
    \midrule
    $q/g$ & native & TTN & 0.8917 $\pm$ 0.0007 & 79.2 & 29.7 & 10.6 \\
    $q/g$ & native & LundNet5 & 0.8941 $\pm$ 0.0007 & 82.1 & 29.9 & 10.9 \\
    $q/g$ & native & ParticleNet & 0.9014 $\pm$ 0.0007 & 89.5 & 33.7 & 12.3 \\
    $q/g$ & native & TTN, gates off & 0.8918 $\pm$ 0.0007 & 79.0 & 29.5 & 10.6 \\
    \addlinespace
    $q/g$ & transfer & TTN & 0.8114 $\pm$ 0.0010 & 38.2 & 12.3 & 4.6 \\
    $q/g$ & transfer & LundNet5 & 0.8098 $\pm$ 0.0010 & 38.2 & 12.4 & 4.6 \\
    $q/g$ & transfer & ParticleNet & 0.8107 $\pm$ 0.0010 & 37.7 & 12.3 & 4.6 \\
    $q/g$ & transfer & TTN, gates off & 0.8124 $\pm$ 0.0010 & 40.0 & 12.6 & 4.6 \\
    \addlinespace
    $t/\mathrm{QCD}$ & native & TTN & 0.9531 $\pm$ 0.0004 & 738.5 & 132.0 & 34.9 \\
    $t/\mathrm{QCD}$ & native & LundNet5 & 0.9574 $\pm$ 0.0004 & 1190.5 & 187.3 & 42.8 \\
    $t/\mathrm{QCD}$ & native & ParticleNet & 0.9592 $\pm$ 0.0004 & 1315.8 & 204.1 & 47.3 \\
    $t/\mathrm{QCD}$ & native & TTN, gates off & 0.9521 $\pm$ 0.0004 & 684.5 & 120.3 & 33.5 \\
    \addlinespace
    $t/\mathrm{QCD}$ & transfer & TTN & 0.9472 $\pm$ 0.0004 & 530.1 & 96.4 & 25.1 \\
    $t/\mathrm{QCD}$ & transfer & LundNet5 & 0.9474 $\pm$ 0.0005 & 769.2 & 125.0 & 29.7 \\
    $t/\mathrm{QCD}$ & transfer & ParticleNet & 0.9479 $\pm$ 0.0005 & 684.9 & 131.9 & 30.4 \\
    $t/\mathrm{QCD}$ & transfer & TTN, gates off & 0.9461 $\pm$ 0.0004 & 495.1 & 95.1 & 25.3 \\
    \bottomrule
  \end{tabular}
  \begin{minipage}{0.98\textwidth}
    \footnotesize The uncertainty beside AUC is DeLong statistical standard error on the test sample; all models are trained with a single seed.  Working points use linear ROC interpolation. 
  \end{minipage}
\end{table*}

\begin{figure*}[htpb]
    \centering
    \includegraphics[width=0.75\textwidth]{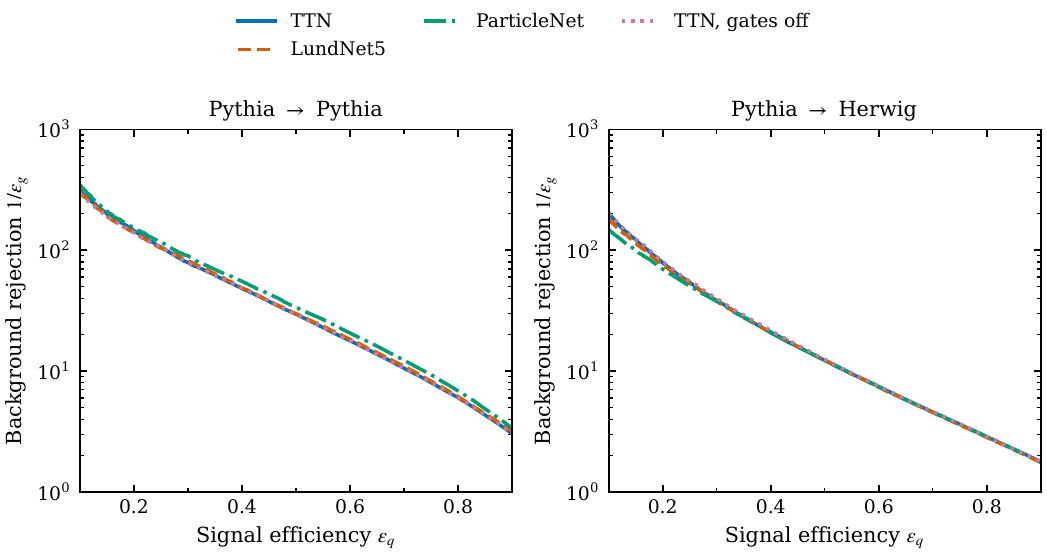}
    \caption{The ROC curves for the quark vs gluons discrimination task are shown, for the taggers trained on Pythia and tested on Pythia (left), trained on Pythia and tested on Herwig (right). The TTN (this work) is shown in blue; LundNet5 in dashed orange and ParticleNet in dash-dotted green lines. TTN is also shown with physics gates off (dotted line). The performance of TTN is competitive across the entire signal efficiency range. }
    \label{fig:qg_baseline}
\end{figure*}
\begin{figure*}[htpb]
    \centering
    \includegraphics[width=0.75\textwidth]{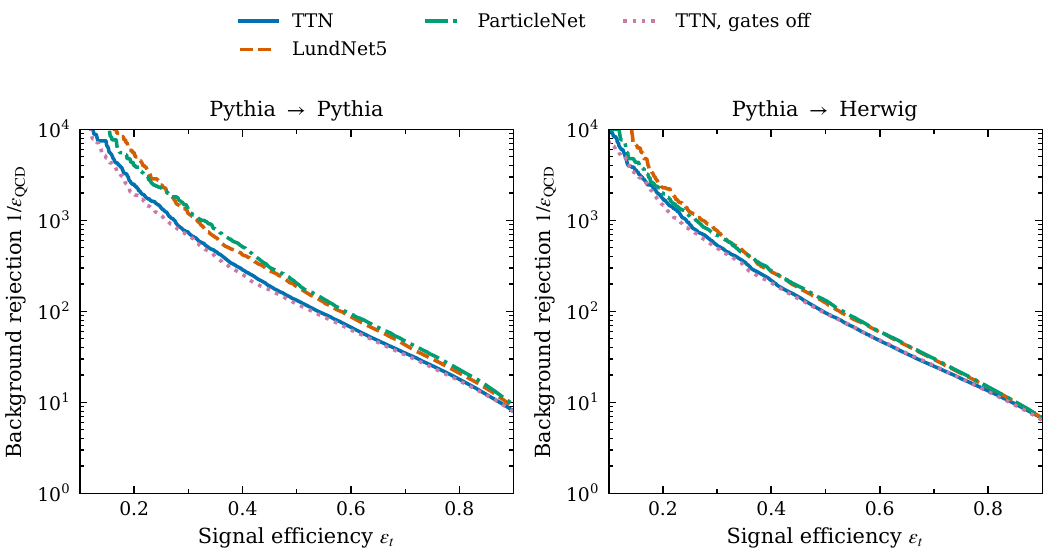}
    \caption{The ROC curves for the top quark vs QCD discrimination task are shown, for the taggers trained on Pythia and tested on Pythia (left), trained on Pythia and tested on Herwig (right). The TTN (this work) is shown in blue; LundNet5 in dashed orange and ParticleNet in dash-dotted green lines. TTN is also shown with physics gates off (dotted line). ParticleNet and LundNet lead in this task; the three taggers show a
    comparable generator dependence.}
    \label{fig:top_baseline}
\end{figure*}

The ROC curves are presented in Figure~\ref{fig:qg_baseline} and Figure~\ref{fig:top_baseline} for the quark vs gluon and top vs QCD discrimination, respectively. We observe that for the first case, a textbook example of a difficult task in terms of out-of-distribution capabilities, the TTN presented in this work achieves a close performance to LundNet and slightly below ParticleNet; additionally, the ROC curves overlap for all of them when the Pythia-trained model is evaluated on Herwig. We also note that both taggers which operate on the Lund plane exhibit a contained degree of degradation if compared to the low-level tagger ParticleNet. 

For the top vs QCD discrimination task, the LundNet and ParticleNet lead the performance across the entire signal efficiency region against the TTN; however when evaluated on the alternative generator, the performances converge, especially in the high signal efficiency region $\epsilon_t>0.8$. 

The relative AUC loss from Pythia to Herwig is 9.0\%, 9.4\% and 10.1\% for
the TTN, LundNet5 and ParticleNet in quark/gluon discrimination, and 0.6\%,
1.0\% and 1.2\% in top tagging. The smaller loss of the TTN partly reflects
its lower nominal performance: the transfer AUCs of the three taggers agree
within 0.002, and in top tagging the TTN retains a lower background rejection than both baselines for $\epsilon_t \lesssim 0.8$.

The dotted curves show the TTN with both physics gates disabled, which
differs from the default by at most 0.0011 in AUC, within about two
standard errors. Since both are evaluated on the same test sample, resolving such differences would require a paired comparison and multiple training seeds; within the present study the gates bring no measurable gain, indicating that the shared core captures the splitting structure without explicit perturbative priors.

\subsection{Capacity scan}\label{subsec:res_2}
In this section we show how the performance varies as a function of the architectural knob of the Tensor Network formulation: the bond dimension $\chi$. This parameter specifies the dimensionality of the recursive tensor which acts on the declustering history. A higher dimensionality allows a higher degree of information to flow upwards from leaf to the root. 

\begin{figure*}[htpb]
    \centering
    \begin{subfigure}[b]{0.48\textwidth}
        \centering
        \includegraphics[width=\textwidth]{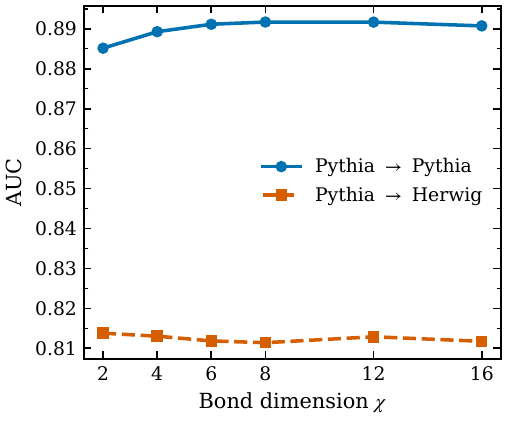}
        \caption{$\chi$ vs AUC for the quark vs gluon discrimination task.}
        \label{fig:scan_chi_qg}
    \end{subfigure}
    \hfill 
    \begin{subfigure}[b]{0.48\textwidth}
        \centering
        \includegraphics[width=\textwidth]{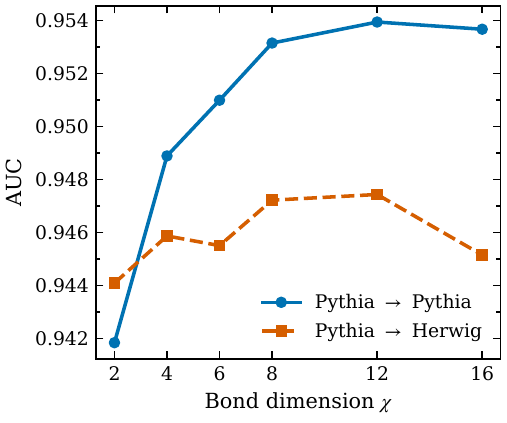}
        \caption{$\chi$ vs AUC for the top vs QCD discrimination task.}
        \label{fig:scan_chi_top}
    \end{subfigure}
    
    \caption{Comparison of AUC vs $\chi$ both tagging tasks. The blue markers show the native (Pythia$\to$Pythia) behavior and the orange markers the transfer (Pythia$\to$Herwig) behavior. For the quark-gluon discrimination, also a low bond dimension allows competitive discrimination. For the complex top quark decay, a higher bond dimension allows carrying information more efficiently across subtrees. }
    \label{fig:scan_chi_combined}
\end{figure*}

In Figure~\ref{fig:scan_chi_combined}, the AUC is reported as a function of the bond dimension. We observe a strikingly different behavior for the two tagging tasks. For the quark vs gluon case, increasing the bond dimension yields only marginal gain in native AUC, while transfer AUC remains constant across the bond-dimension space when evaluated on Herwig. We interpret this in terms of the QCD cascade: quark and gluons produce quarks and other gluons, so the shared tensor core is already well suited because the nature of the interaction is governed by the same principle in each iteration.
For the top tagging task instead we observe a different behavior: the increase in native AUC is more important, and slightly correlates with the transfer one. In this case the top quark decays with a complex topology, meaning that separated subtrees representing the three prongs are more strongly correlated than individual subtrees in quark- and gluon-jets. For this reason a higher bond dimension helps avoiding information bottleneck at each node.

\subsection{$\ln(k_t/\text{GeV})$ threshold scan}
\label{subsec:res_3}

\begin{figure*}[htpb]
    \centering
    \begin{subfigure}[b]{0.48\textwidth}
        \centering
        \includegraphics[trim=0mm  7mm 0mm 0mm, clip, width=\textwidth]{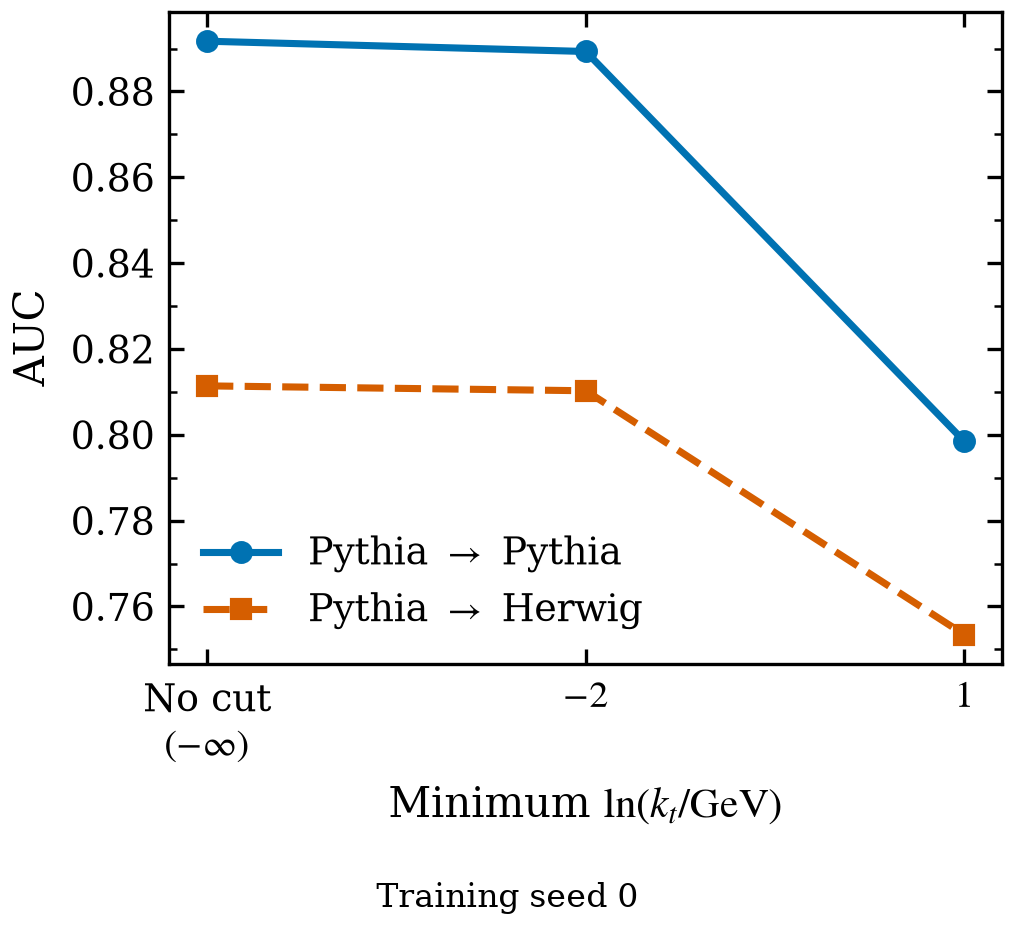}
        \caption{$\ln(k_t/\text{GeV})$ vs AUC for the quark vs gluon discrimination task.}
        \label{fig:scan_chi_qg}
    \end{subfigure}
    \hfill 
    \begin{subfigure}[b]{0.48\textwidth}
        \centering
        \includegraphics[trim=0mm  7mm 0mm 0mm, clip, width=\textwidth]{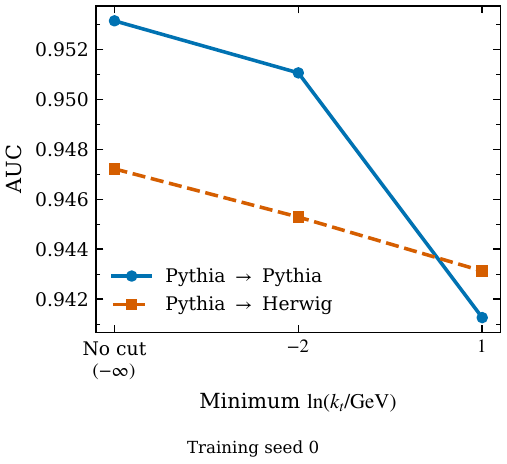}
        \caption{$\ln(k_t/\text{GeV})$ vs AUC for the top vs QCD discrimination task.}
        \label{fig:auc_vs_log_kt_cut}
    \end{subfigure}
    
    \caption{Comparison of AUC vs $\ln(k_t/\text{GeV})$  both tagging tasks. The blue markers show the native (Pythia$\to$Pythia) behavior and the orange markers the transfer (Pythia$\to$Herwig) behavior. For the quark-gluon discrimination, there is no decrease of performance up to $\ln(k_t/\text{GeV})=-2$, while the relative decrease from -2 to 1 is stronger in the native with respect to the transfer (Pythia$\to$Herwig) case. For the top quark, the decrease at -2 is of similar size (about 0.002), while at $\ln(k_t/\text{GeV})=1$ the transfer AUC exceeds the native one.}
    \label{fig:scan_logkt_combined}
\end{figure*}

In this Subsection, the perturbative threshold is $\ln(k_t/\text{GeV})$ varied from $-\infty$ (no cut) to -2 and 1; conventionally values above 0 are phenomenologically interpreted as above the perturbativity scale. Effectively, in the declustering tree, splittings below the threshold are skipped by following the harder branch (Section~\ref{sec:ttnmethod}), removing soft, non-perturbative emissions. This robustness study was already performed for LundNet~\cite{Dreyer:2020brq}, and we reproduce it here for the TTN. This control of perturbativity has no counterpart in unconstrained constituent level taggers such as ParticleNet.
In Figure~\ref{fig:scan_logkt_combined} we observe a negligible decrease in performance for both native and transfer regimes for the quark/gluon tagging up to $\ln(k_t/\text{GeV})=-2$. At this cut, where the classifier is IRC safe, the native AUC decreases by at most about 0.003 in both tasks. Instead, for the top quark, the transfer case exhibits a smaller degree of degradation, inverting the order at $\ln(k_t/\text{GeV})=1$, where the Pythia$\to$Herwig case shows a higher AUC. A similar inversion was also found earlier in the case of $\chi=2$. 

\subsection{Identifying the latent representation changes}\label{subsec:res_4}
In this Section, we show a distinctive way to employ the TTN approach presented in this paper. As detailed in the architecture Section~\ref{sec:ttnmethod}, the tensor core is identically applied recursively along the shower history. This is very different from graph-based large neural networks, which construct edge convolutions along each node connection.
We use this construction in the following way to learn more about the jet and its different description in the two generators. 

One interesting question in this regard is where in the tree depth the discrimination arises. To answer this, we evaluate the fully trained TTN on truncated trees, starting from root, and replace truncated nodes with the learned vacuum state. We emphasize that this is better defined for the TTN than for other approaches based on the declustering history of the jet: in the TTN a truncated subtree is replaced by $h_\varnothing$, the same state the model uses at every leaf in training, so the truncated jet is processed by exactly the same bottom-up map. In graph neural networks, even with weights shared across nodes as in EdgeConv, truncation changes the graph on which message passing is performed: neighborhoods, paths and the set of states entering the global aggregation are modified.
The model is trained on full trees only, so the absolute AUC at small depth
also reflects the response of the readout to inputs it has not been trained
on; we therefore focus on the Pythia/Herwig ratio, where this effect is
largely common to both generators.

To characterize the individual classes, we utilize the mean predicted positive-class probability instead of AUC, $p_D=1/(1+e^{-s_D})$, where $D$ is the tree depth, and $s_D$ is the classifier score obtained after truncating the tree at depth $D$. Additionally, we show the cosine distance between root $h_{root}$ for the tree truncated at $D$ with respect to the one truncated at $D-1$. This measures how much the final root representation changes when one additional branching level is exposed: a value near zero means that adding the next branching barely changes the root representation; conversely large values imply a large impact at root level. The cosine distance is defined as usual:
\[
1-\cos\!\left(h_{\mathrm{root}}^{(D)},\,h_{\mathrm{root}}^{(D-1)}\right)
=
1-
\frac{
h_{\mathrm{root}}^{(D)} \cdot h_{\mathrm{root}}^{(D-1)}
}{
\left\|h_{\mathrm{root}}^{(D)}\right\|
\left\|h_{\mathrm{root}}^{(D-1)}\right\|
}.
\]

\begin{figure*}[htpb]
    \centering
    
    \begin{subfigure}[b]{0.45\textwidth}
        \centering
        \includegraphics[width=\linewidth]{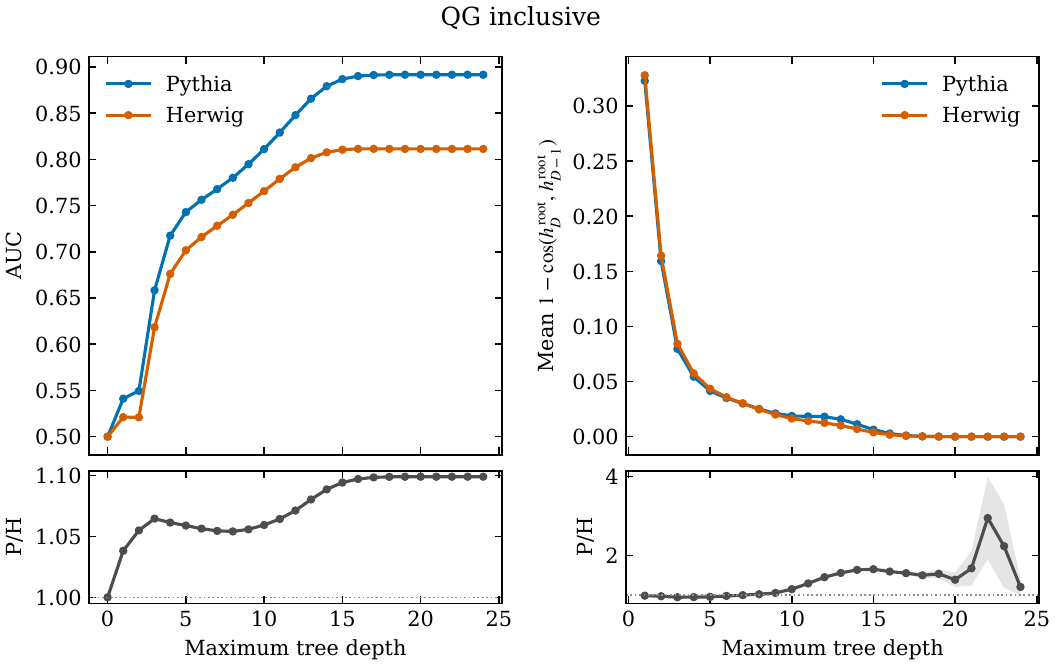}
        \caption{AUC and $h_{\mathrm{root}}$ cosine distance vs tree depth for quark and gluon discrimination task.}
        \label{fig:qg_depth_overall}
    \end{subfigure}
    \hfill
    \begin{subfigure}[b]{0.45\textwidth}
        \centering
        \includegraphics[width=\linewidth]{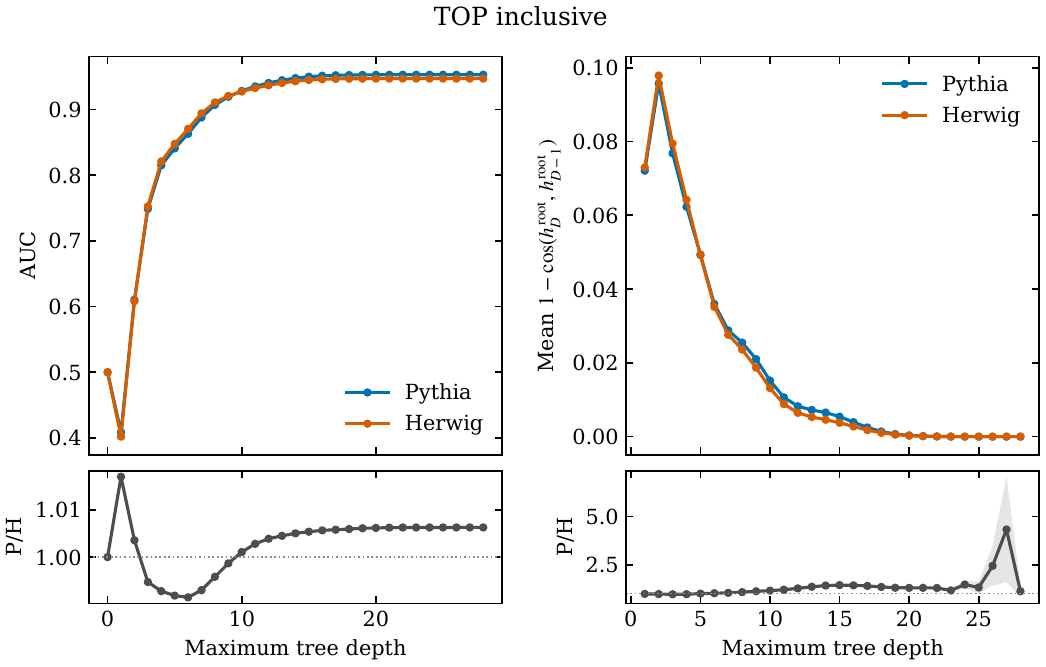}
        \caption{AUC and $h_{\mathrm{root}}$ cosine distance vs tree depth for top vs QCD discrimination task.}
        \label{fig:top_depth_overall}
    \end{subfigure}

    \caption{AUC and $h_{\mathrm{root}}$ cosine distance for quark vs gluon (left) and top vs QCD (right) tasks as a function of tree depth. }
    \label{fig:combined_AUC_diagnostics}
\end{figure*}

In Figure~\ref{fig:combined_AUC_diagnostics}, we show global AUC and the $h_{\rm root}$ cosine distance for Pythia (native) and Herwig (transfer) as a function of the tree depth, for quark and gluons (left) and top quark and QCD (right). The complete breakdown for the different classes individually is shown in Figure~\ref{fig:combined_all_diagnostics} in ~\ref{sec:appA}.

We observe the following: in the quark vs gluon tagging task, the AUC accumulates fast for the first 5 declustering levels; changing regime until around a depth of 15 and then saturates. The difference with Herwig suggests differences growing up constantly from around depth 3; while the cosine difference shows different behavior between tree level 10-15, which correspond to the core of the small-angle soft-emitting regime. Looking separately at quarks and gluons, we observe a higher degree of dissimilarity between Pythia and Herwig for gluons, in the region depth $>5$, with strong contribution from the region 10-15.
For top tagging task, the behavior is different: the AUC increases fast for the first five branching, then slows down and saturates at around 15, with Pythia to Herwig differences accumulating in different directions. Inspecting the QCD class for top tagging, we recover a similar behavior as in the quark vs gluon task.
For top quarks, the mean probability moves in the opposite direction, again getting differences especially in the 10-15 region.

\begin{figure*}[htpb]
    \centering
    \includegraphics[width=0.70\textwidth]{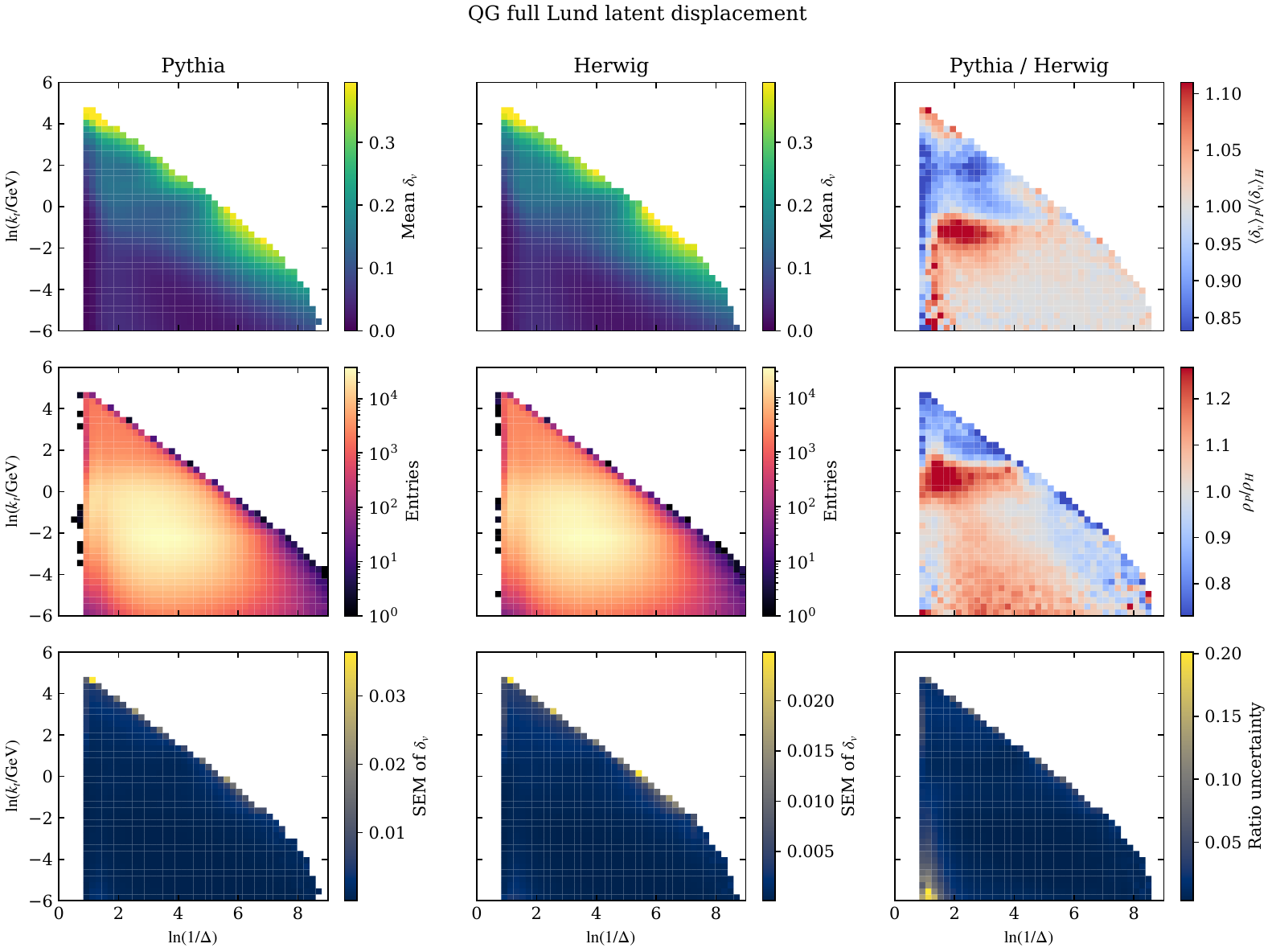}
    \caption{Top row: mean cosine distance of the latent state with its hardest child, $\delta_v$ for Pythia (left), Herwig (center) and their ratio (right), for the quark-gluon task in the full Lund plane. Central row: Lund density for Pythia, Herwig and their ratio. Bottom row: standard error on the mean of $\delta_v$ for Pythia, Herwig and their ratio.}
    \label{fig:qg_full_density}
\end{figure*}

\begin{figure*}[htpb]
    \centering
    \includegraphics[width=0.70\textwidth]{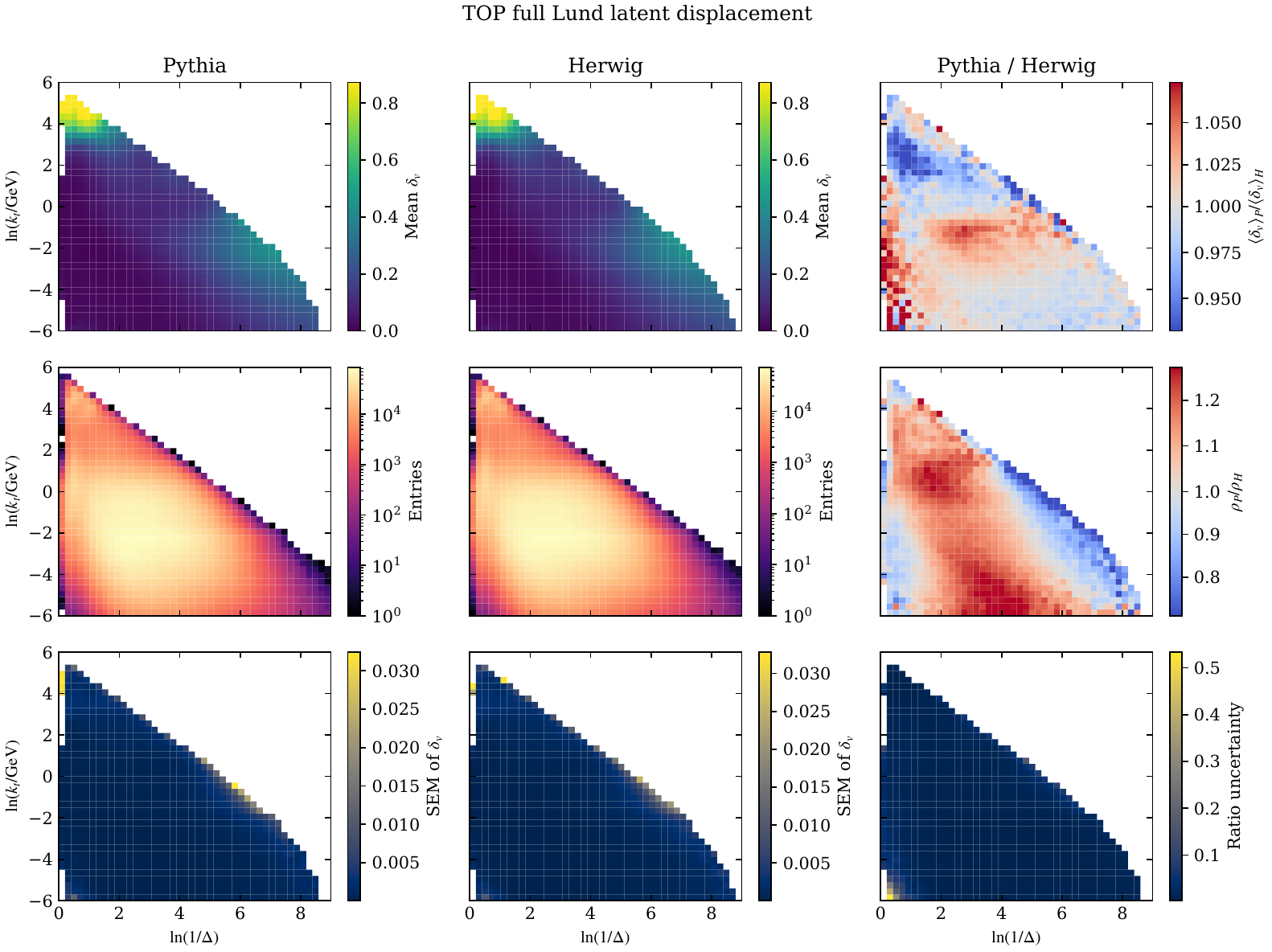}
    \caption{Top row: mean cosine distance of the latent state with its hardest child, $\delta_v$ for Pythia (left), Herwig (center) and their ratio (right), for the top tagging task in the full Lund plane. Central row: Lund density for Pythia, Herwig and their ratio. Bottom row: standard error on the mean of $\delta_v$ for Pythia, Herwig and their ratio.}
    \label{fig:topqcd_full_density}
\end{figure*}

We study the cosine distance of $h_v$ with its hardest child, and call this quantity $\delta_v$. Also in this case, the interpretation is the following: how much the latent representation changes going upstream in a splitting node with respect to the hardest child. This allows us to understand exactly where in the Lund plane the representation is changing more significantly, and to compare if this behavior is different in Pythia and Herwig and exactly where. We compare this measure with the nominal Lund plane density, for the full and primary Lund plane. 

For the quark gluon tagging dataset, we observe that the most of the cosine difference accumulates along the kinematic border, while the rest of the plane does show smaller degree of difference. This means that the shared tensor core changes the latent representation by a high degree when processing the hard child which happens to have high $\ln(k_t/\text{GeV})$ or very collinear. Interestingly, the Lund plane differs in Pythia with respect to Herwig mostly at the onset of the non perturbative region $\ln(k_t/\text{GeV}) < 1$, while for $h_v$ the differences concentrate mostly at $\ln(k_t/\text{GeV})<0$. These effects are visible in Figure~\ref{fig:qg_full_density}; in the primary Lund plane, Figure~\ref{fig:qg_primary_density}, the differences move to the $\ln(1/\Delta)>4$ and $\ln(k_t/\text{GeV})<0$ instead, the softer collinear region.

In the top tagging dataset, we observe that the highest values of $\delta_v$ are concentrated in the top region of the Lund plane, the hard emissions at large angle. This is expected: the majority of the splittings in the declustering tree obey the showering and hadronization of quarks and gluons; however their origin is from the electroweak decay of the top quark, therefore the first hard splittings are related to b-quark and $W$ boson branching and its decay into quarks. In this sense the latent representation finds this region more distant in behavior with respect to the rest of the jet. Again the region $\ln(k_t/\text{GeV})<0$ and interestingly a narrow region at large emission accumulate the most distance in latent representation change. Also in this case, these effects are visible in Figure~\ref{fig:topqcd_full_density}; in the primary Lund plane, Figure~\ref{fig:topqcd_primary_density}, the differences move to the $\ln(1/\Delta)>4$ and $\ln(k_t/\text{GeV})<0$ instead, the softer collinear region.
A breakdown of this Lund plane displacement is presented also for the individual classes in~\ref{sec:appA}.

\section{Conclusions}
\label{sec:conlusions}
In this work, we have presented a novel construction for jet tagging on the Lund declustering tree in which IRC-safety is controlled through a low $k_t$ cut on the tree. Using a quantum-inspired Tree Tensor Network approach, where a shared low-bond-dimension tensor core is applied recursively across each node, we achieve performance close to LundNet and ParticleNet in quark/gluon tagging, and a slightly lower background rejection in top tagging, with a Pythia-to-Herwig transfer AUC within 0.002 of both baselines at about 1\% of the parameters of LundNet. With a cut at $\ln(k_t/\text{GeV})=-2$ the classifier is IRC safe at a cost of at most about 0.003 in AUC, and raising the cut toward the perturbative scale reduces the Pythia$\to$Herwig gap at the price of discrimination.
The recursive structure and the shared core allow us to precisely identify changes of the latent configuration on the tree depth and on the Lund plane, qualifying additionally where Pythia and Herwig differ the most in the latent space representation. This potentially allows the TTN to be trained and evaluated on regions of the phase space where the theoretical uncertainties on the showering and hadronization model are less severe. 
Finally, ATLAS, CMS and LHCb~\cite{ATLAS:2020bbn,CMS:2023lpp,LHCb:2025mcq} have measured and unfolded the primary Lund jet plane and its uncertainty; the regions on which the TTN relies could be chosen where experimental and modelling uncertainties are smallest; for low-level taggers based on large networks, these uncertainties can be substantial~\cite{SciPostPhys.17.6.166}. Reducing this part of the uncertainty budget while maintaining high efficiency will be of great relevance at HL-LHC, where statistically limited measurements and searches will be forced to carefully consider the tagging strategy and its impact to the final sensitivity. A full assessment of generator robustness will require efficiencies at fixed classifier threshold, the complete Pythia/Herwig training–evaluation matrix, multiple training seeds, and parameter-matched baselines. We leave this for future work with the TTN approach.

\begin{acknowledgements}
The authors gratefully acknowledge the UniNuvola Cloud infrastructure of the University of Perugia and INFN for providing the high-performance computing resources used in this study.

This work was funded by the project “BOOST – Boosted Object and Oriented-Space Topologies from VBS@HL-LHC” (CUP I57G21000110007), financed through the cascade call of SPOKE 2 – “Fundamental Research \& Space Economy” of the PNRR ICSC project (code CN00000013, CUP I53C21000340006), Missione 4 – Componente 2 – Investimento 1.4 “Potenziamento strutture di ricerca e creazione di ‘campioni nazionali di R\&S’ su alcune Key Enabling Technologies”. This work was also supported by the project “Quanta – Laboratorio di calcolo quantistico” of the University of Perugia.
\end{acknowledgements}

\appendix
\section{Cosine distance of the latent representation across the Lund Plane}
\label{sec:appA}
In this appendix we report the cosine distance of the latent representation for the full and primary Lund plane, for both tasks, together per individual classes.

\begin{figure*}[htpb]
    \centering
    
    \begin{subfigure}[b]{0.48\textwidth}
        \centering
        \includegraphics[width=\linewidth]{figures/qg_baseline_latent_diagnostics_v4/depth_diagnostics.pdf}
        \caption{AUC and $h_{\mathrm{root}}$ cosine distance vs tree depth for quark and gluon discrimination task.}
        \label{fig:qg_depth_overall}
    \end{subfigure}
    \hfill
    \begin{subfigure}[b]{0.48\textwidth}
        \centering
        \includegraphics[width=\linewidth]{figures/top_baseline_latent_diagnostics_v4/depth_diagnostics.pdf}
        \caption{AUC and $h_{\mathrm{root}}$ cosine distance vs tree depth for top vs QCD discrimination task.}
        \label{fig:top_depth_overall}
    \end{subfigure}
    
    \vspace{0.4cm}
    
    \begin{subfigure}[b]{0.48\textwidth}
        \centering
        \includegraphics[width=\linewidth]{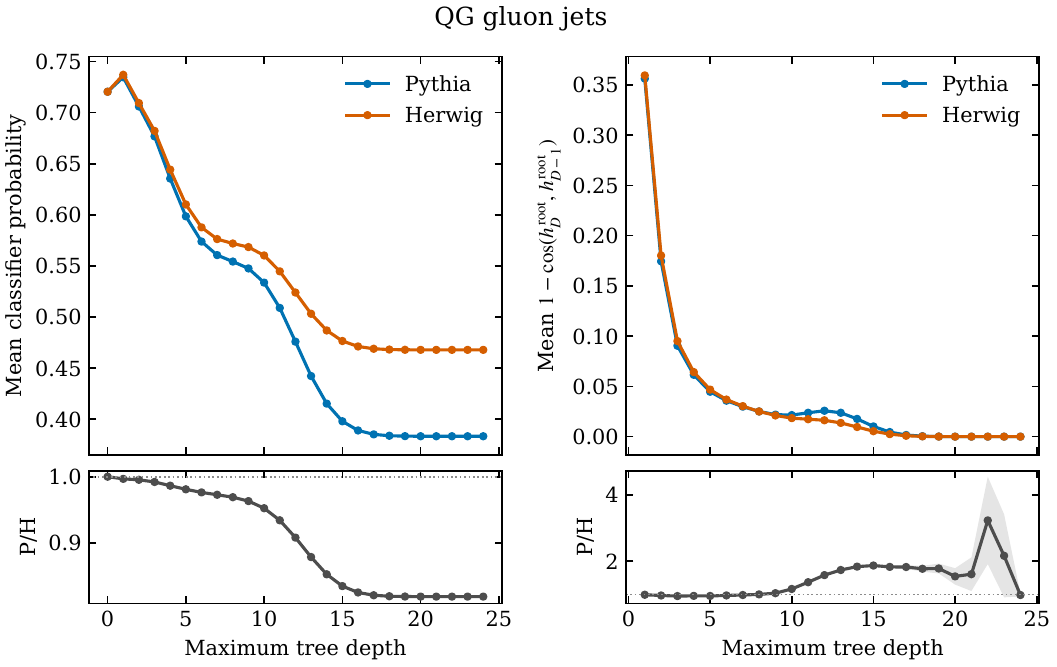}
        \caption{Classifier probability $p_D$ and $h_{\mathrm{root}}$ cosine distance vs tree depth for gluon-jets.}
        \label{fig:qg_depth_gluon}
    \end{subfigure}
    \hfill
    \begin{subfigure}[b]{0.48\textwidth}
        \centering
        \includegraphics[width=\linewidth]{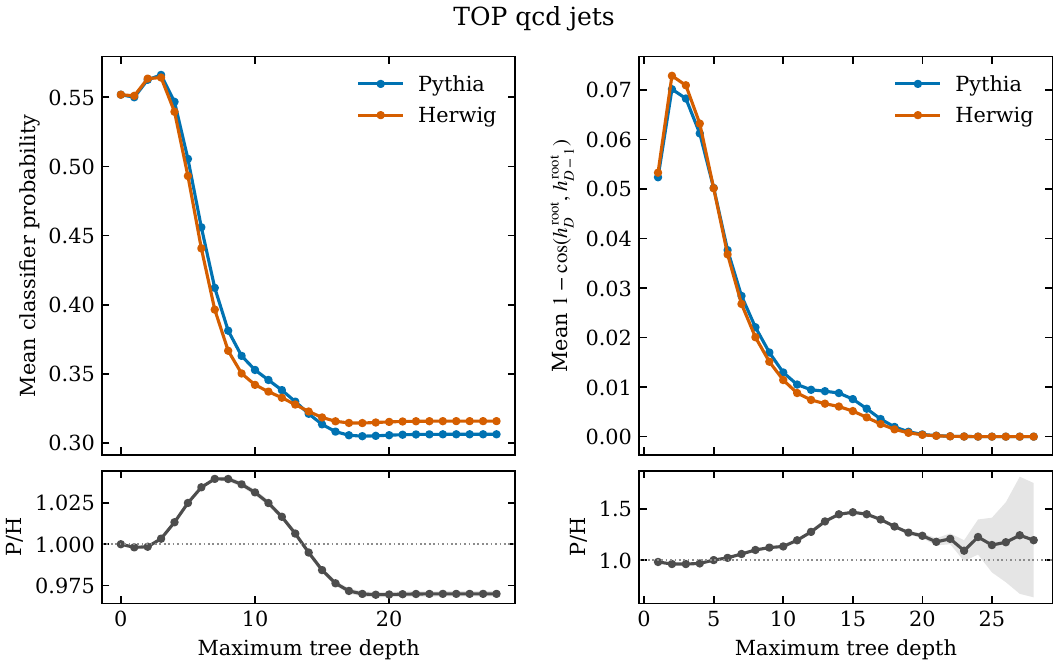}
        \caption{Classifier probability $p_D$ and $h_{\mathrm{root}}$ cosine distance vs tree depth for QCD-jets.}
        \label{fig:top_depth_qcd}
    \end{subfigure}
    
    \vspace{0.4cm}
    
    \begin{subfigure}[b]{0.48\textwidth}
        \centering
        \includegraphics[width=\linewidth]{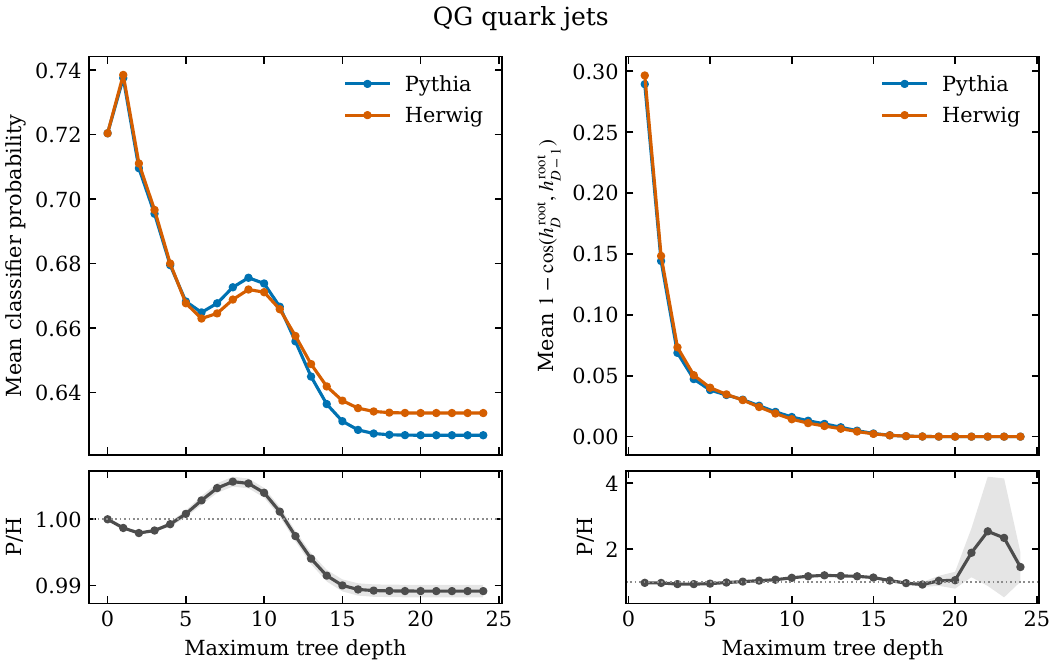}
        \caption{Classifier probability $p_D$ and $h_{\mathrm{root}}$ cosine distance vs tree depth for quark-jets.}
        \label{fig:qg_depth_quark}
    \end{subfigure}
    \hfill
    \begin{subfigure}[b]{0.48\textwidth}
        \centering
        \includegraphics[width=\linewidth]{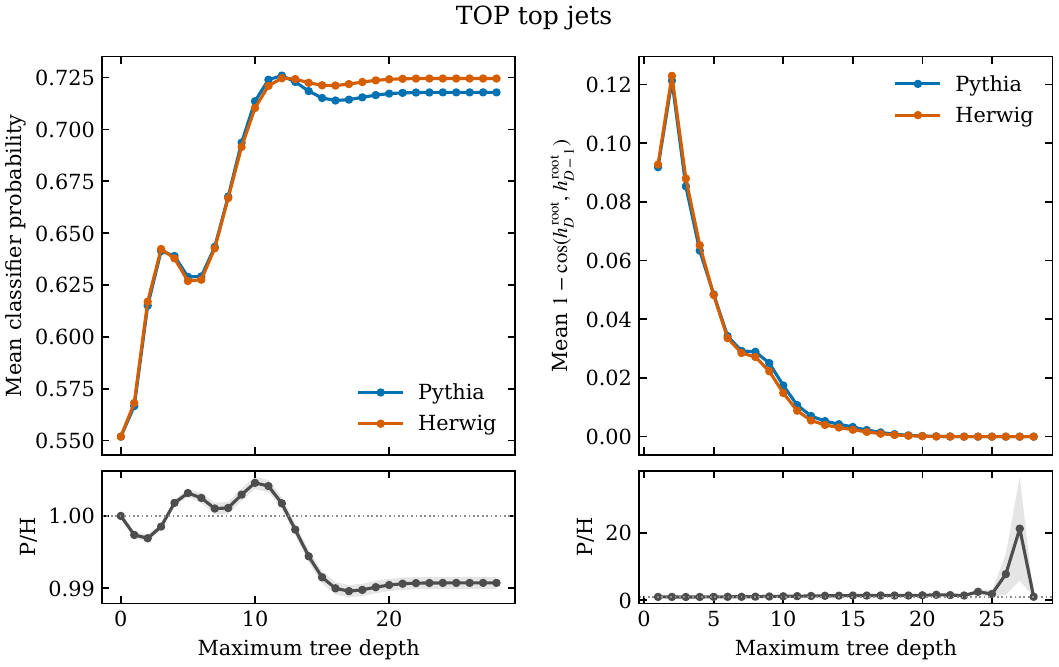}
        \caption{Classifier probability $p_D$ and $h_{\mathrm{root}}$ cosine distance vs tree depth for top-jets.}
        \label{fig:top_depth_top}
    \end{subfigure}
    
    \caption{Comparison for quark vs gluon (left column) and top vs QCD (right column) tasks as a function of tree depth. The first row shows the inclusive results, the second and third rows the individual classes. The comparison is made in terms of AUC and $h_{\mathrm{root}}$ cosine distance (top row), and  $p_D$ and $h_{\mathrm{root}}$ cosine distance (central and bottom row)}
    \label{fig:combined_all_diagnostics}
\end{figure*}

\begin{figure*}[htpb]
    \centering
    \includegraphics[width=0.70\textwidth]{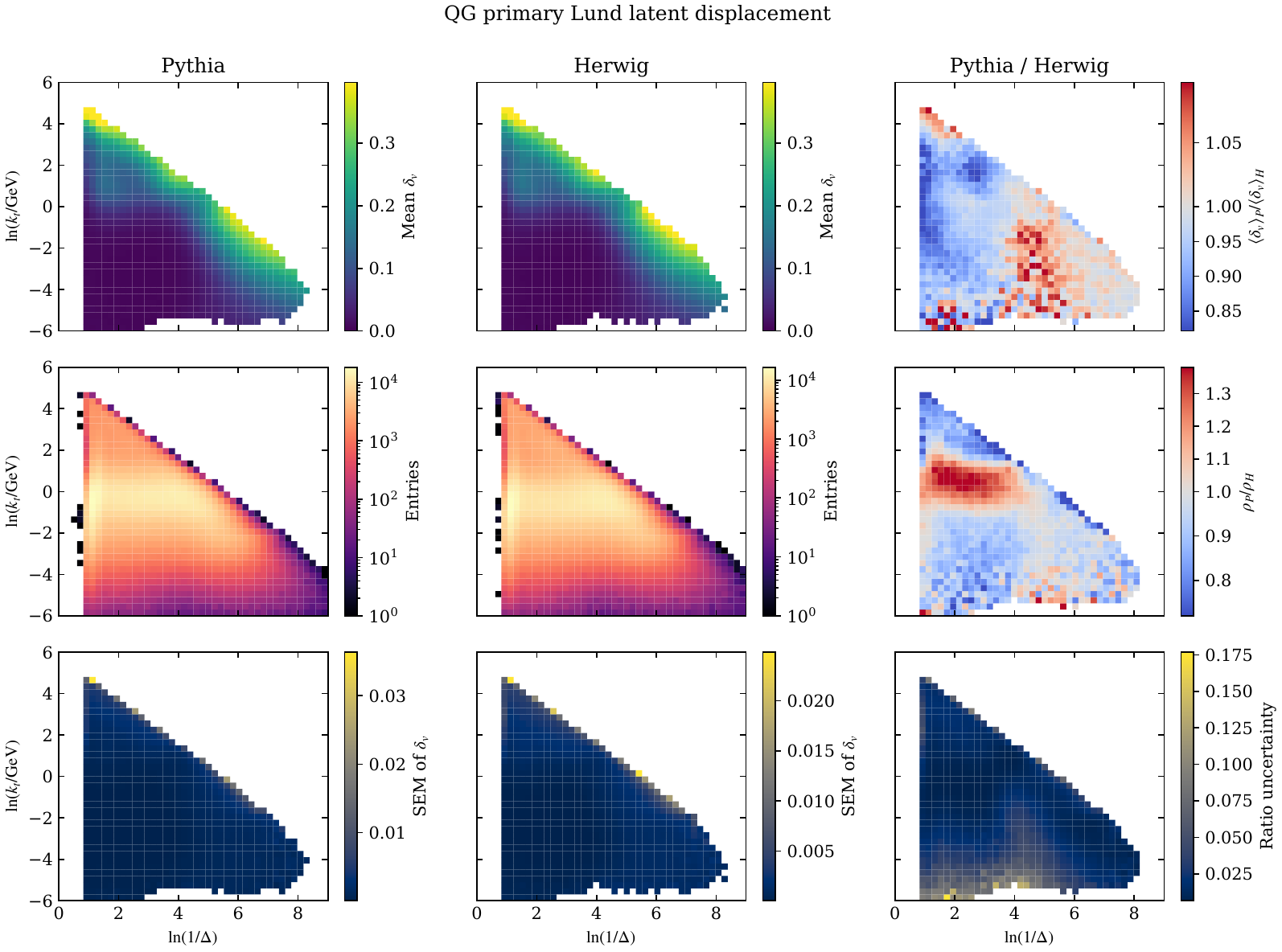}
    \caption{Top row: mean cosine distance of the latent state with its hardest child, $\delta_v$ for Pythia (left), Herwig (center) and their ratio (right), for the quark-gluon task in the primary Lund plane. Central row: primary Lund density for Pythia, Herwig and their ratio. Bottom row: standard error on the mean of $\delta_v$ for Pythia, Herwig and their ratio.}
    \label{fig:qg_primary_density}
\end{figure*}

\begin{figure*}[htpb]
    \centering
    \includegraphics[width=0.70\textwidth]{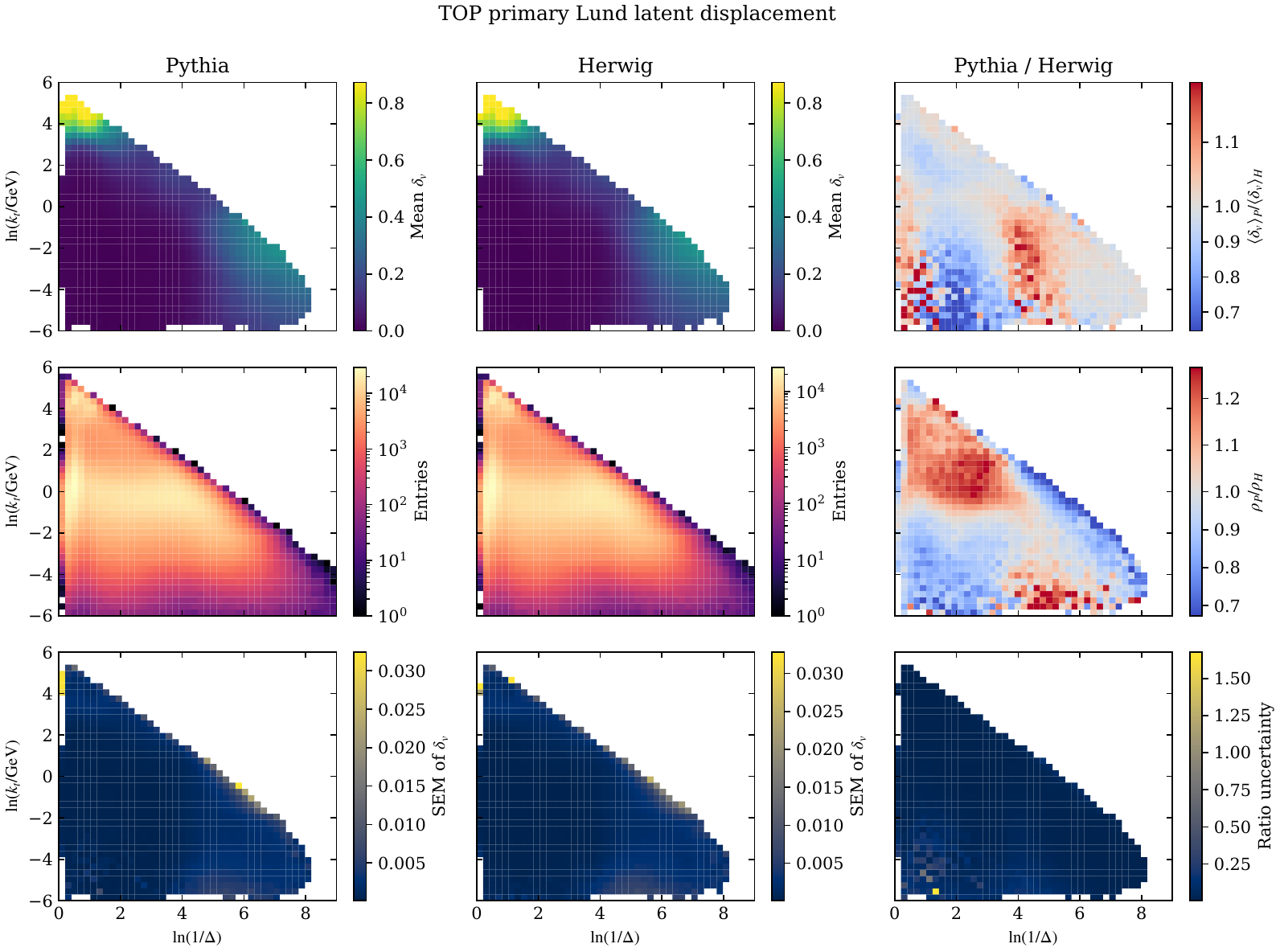}
    \caption{Top row: mean cosine distance of the latent state with its hardest child, $\delta_v$ for Pythia (left), Herwig (center) and their ratio (right), for the top tagging task in the primary Lund plane. Central row: primary Lund density for Pythia, Herwig and their ratio. Bottom row: standard error on the mean of $\delta_v$ for Pythia, Herwig and their ratio.}
    \label{fig:topqcd_primary_density}
\end{figure*}

\begin{figure*}[htpb]
    \centering
    \includegraphics[width=0.70\textwidth]{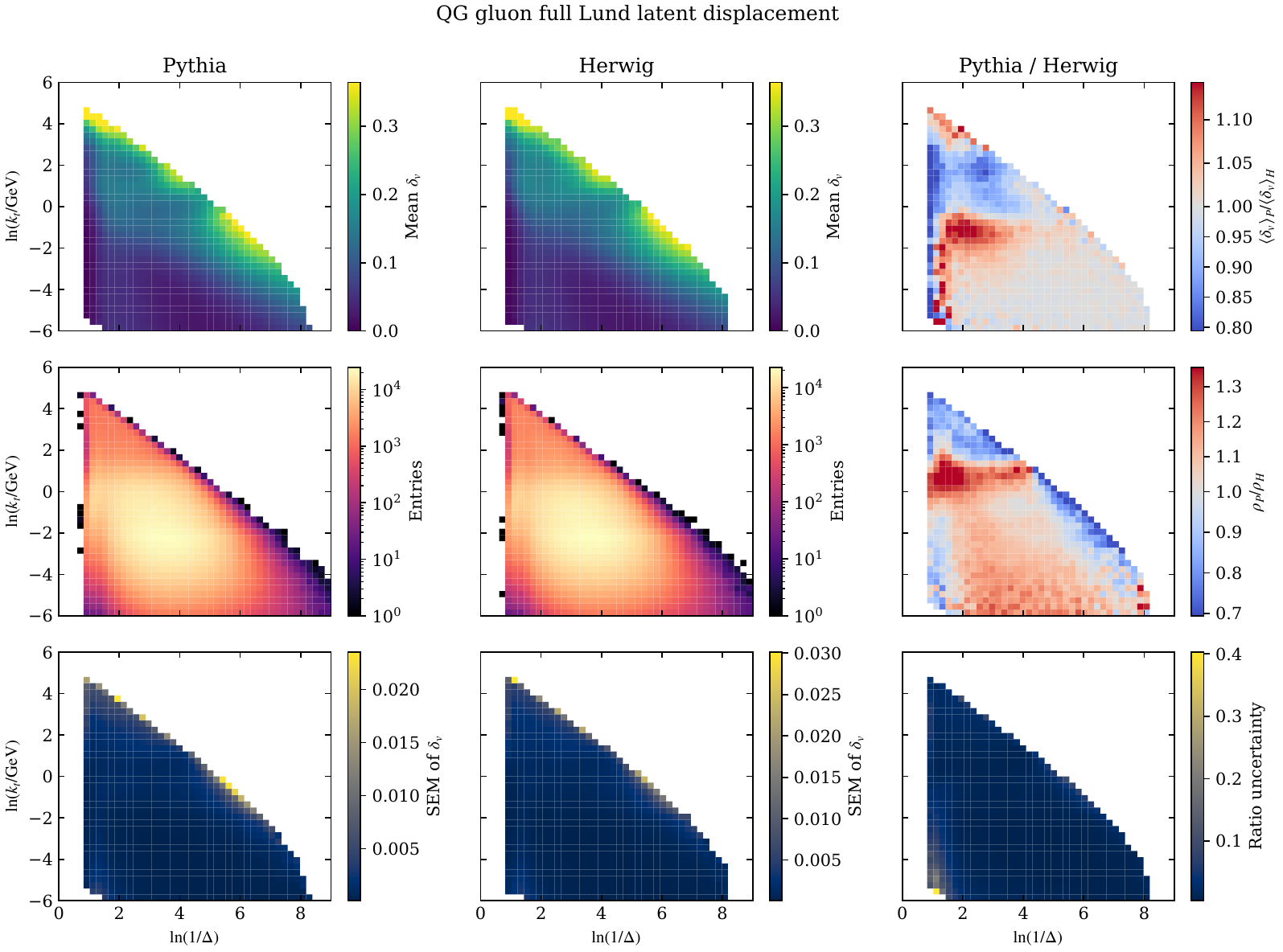}
    \caption{Top row: mean cosine distance of the latent state with its hardest child, $\delta_v$ for Pythia (left), Herwig (center) and their ratio (right), for the gluon class in the quark-gluon task (full Lund plane). Central row: Lund density for Pythia, Herwig and their ratio. Bottom row: standard error on the mean of $\delta_v$ for Pythia, Herwig and their ratio.}
    \label{fig:g_full_density}
\end{figure*}

\begin{figure*}[htpb]
    \centering
    \includegraphics[width=0.70\textwidth]{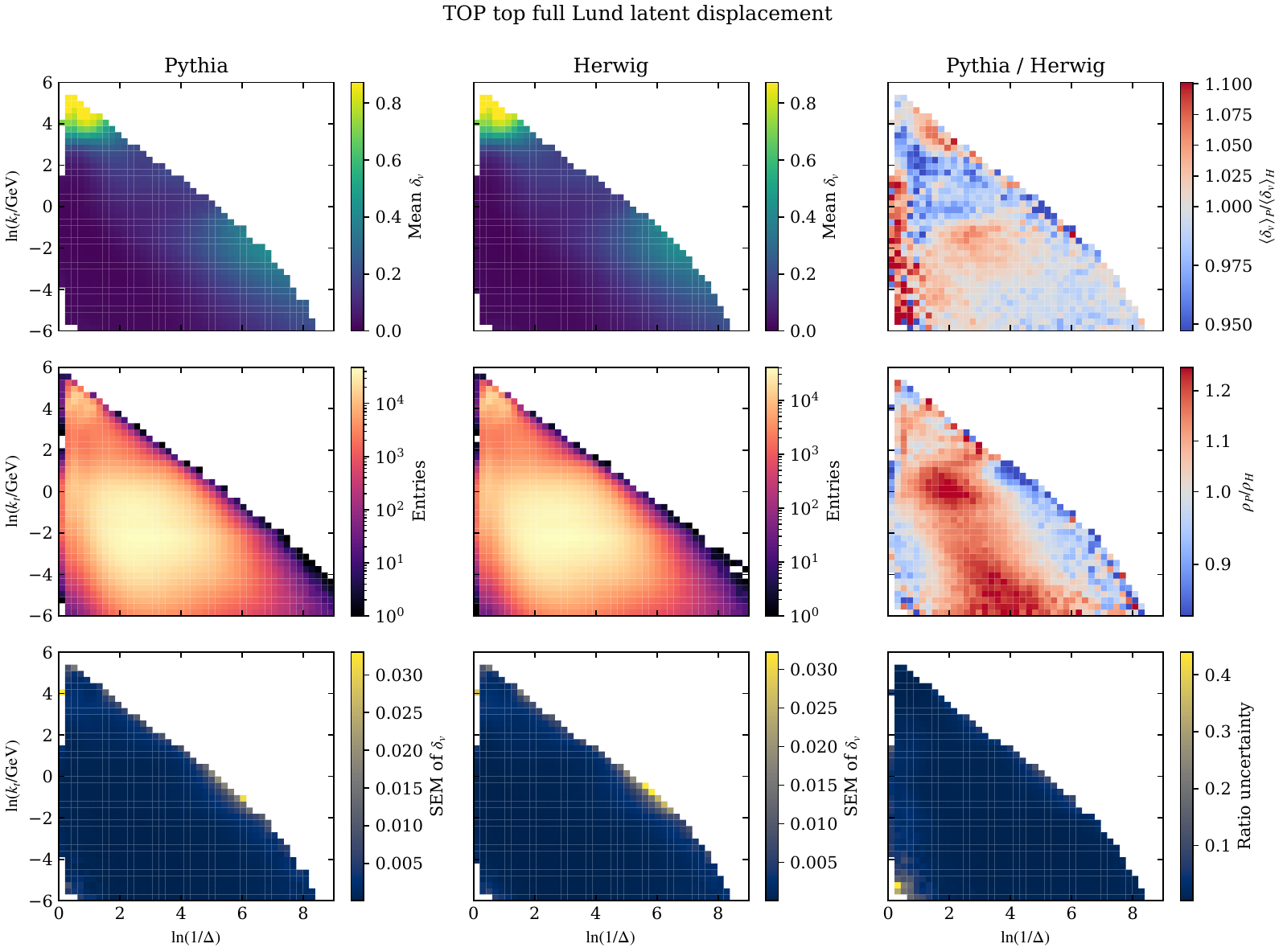}
    \caption{Top row: mean cosine distance of the latent state with its hardest child, $\delta_v$ for Pythia (left), Herwig (center) and their ratio (right), for the top class in the top tagging task (full Lund plane). Central row: Lund density for Pythia, Herwig and their ratio. Bottom row: standard error on the mean of $\delta_v$ for Pythia, Herwig and their ratio. }
    \label{fig:top_full_density}
\end{figure*}

\begin{figure*}[htpb]
    \centering
    \includegraphics[width=0.70\textwidth]{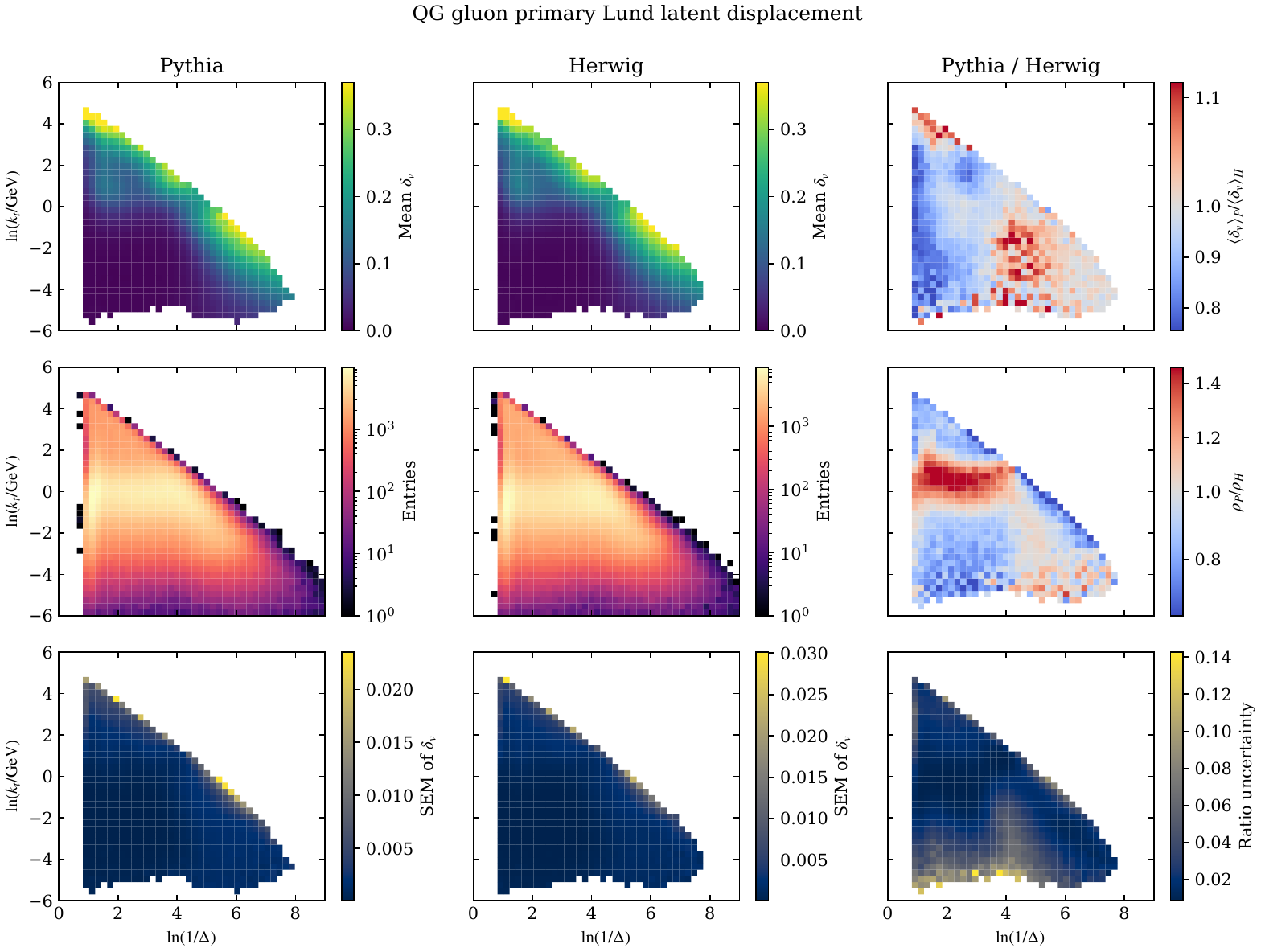}
    \caption{Top row: mean cosine distance of the latent state with its hardest child, $\delta_v$ for Pythia (left), Herwig (center) and their ratio (right), for the gluon class in the quark-gluon task (primary Lund plane). Central row: primary Lund density for Pythia, Herwig and their ratio. Bottom row: standard error on the mean of $\delta_v$ for Pythia, Herwig and their ratio.}
    \label{fig:g_primary_density}
\end{figure*}

\begin{figure*}[htpb]
    \centering
    \includegraphics[width=0.70\textwidth]{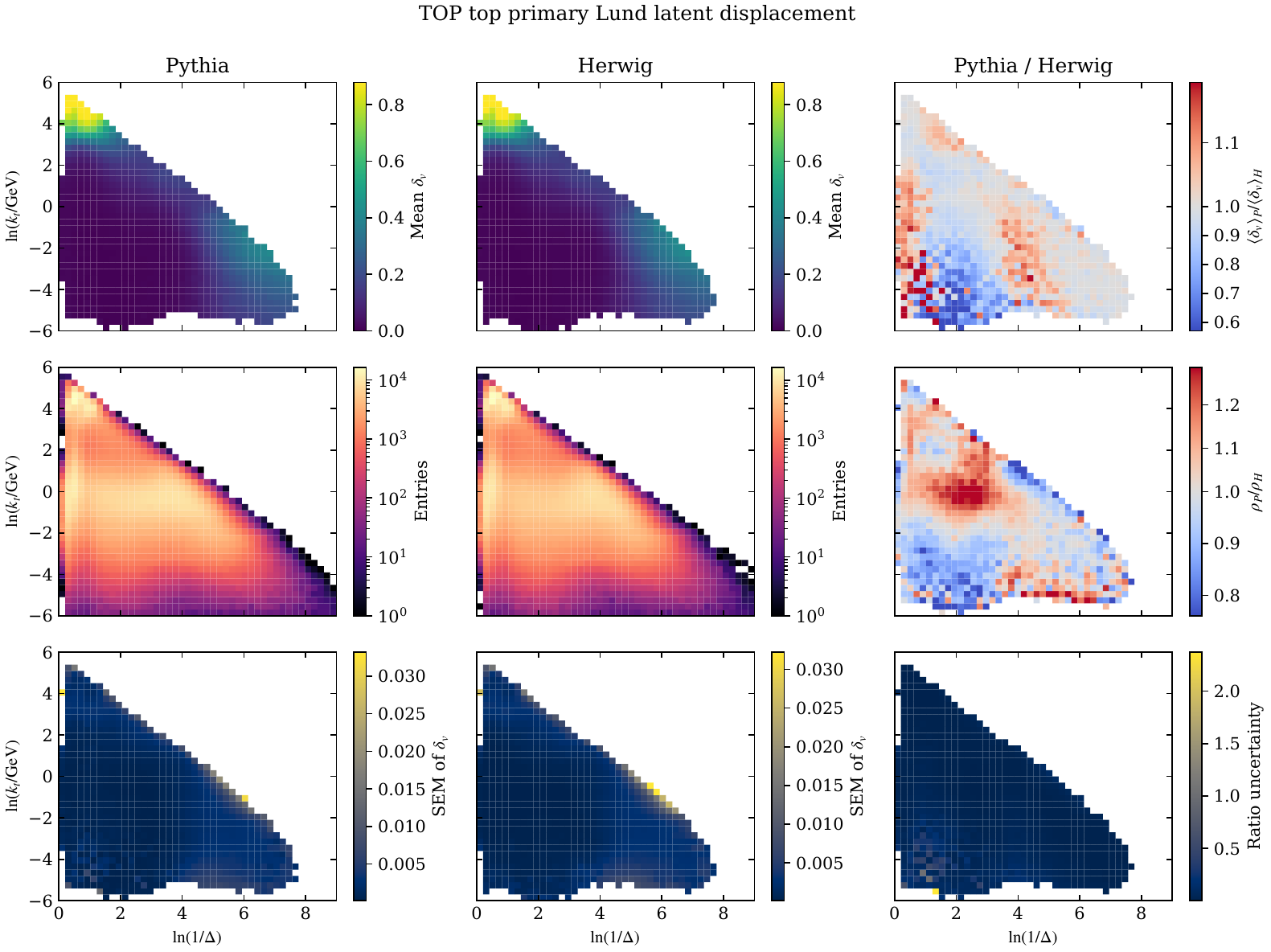}
    \caption{Top row: mean cosine distance of the latent state with its hardest child, $\delta_v$ for Pythia (left), Herwig (center) and their ratio (right), for the top class in the top tagging task (primary Lund plane). Central row: primary Lund density for Pythia, Herwig and their ratio. Bottom row: standard error on the mean of $\delta_v$ for Pythia, Herwig and their ratio.}
    \label{fig:top_primary_density}
\end{figure*}

\begin{figure*}[htpb]
    \centering
    \includegraphics[width=0.70\textwidth]{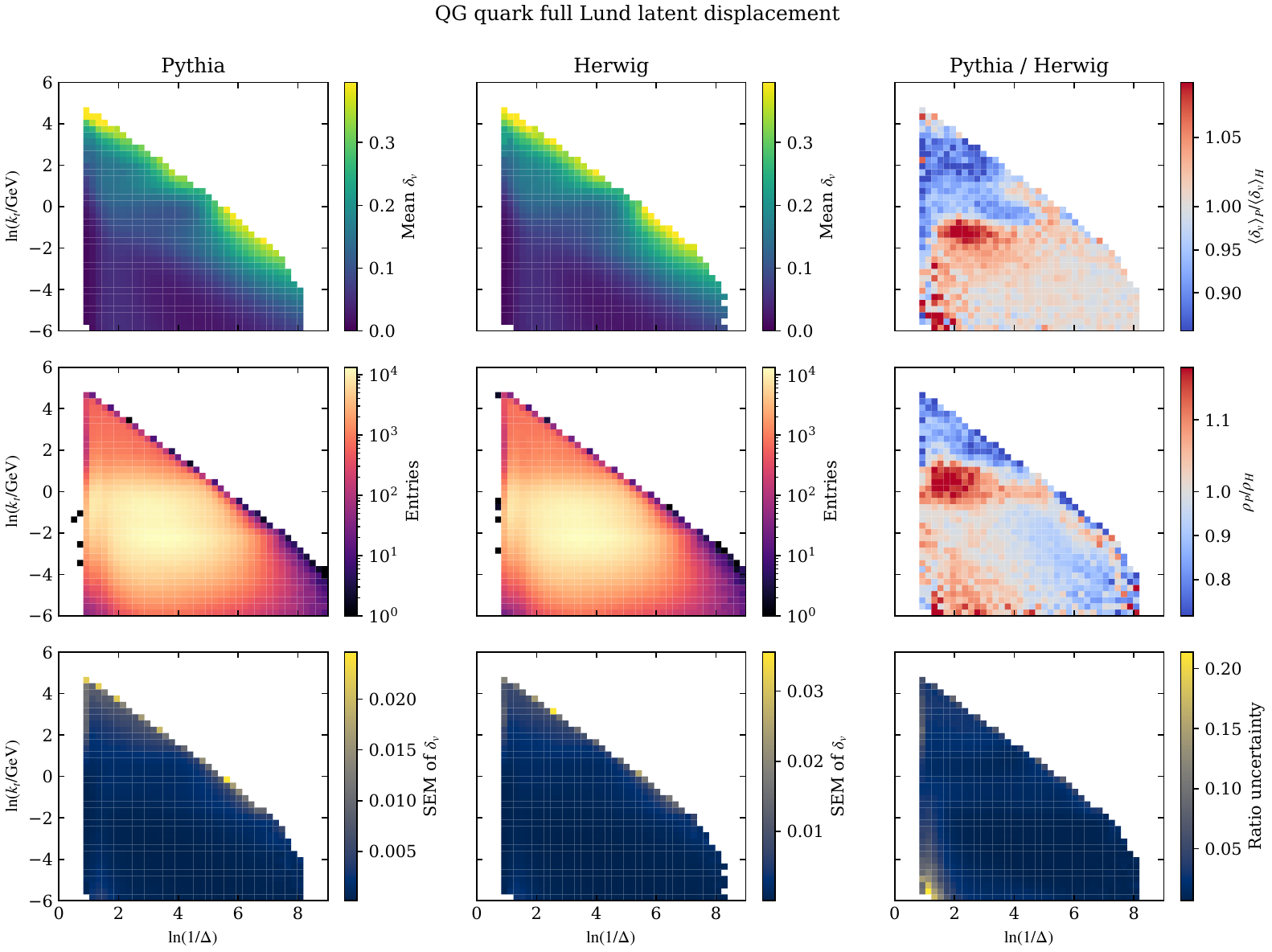}
    \caption{Top row: mean cosine distance of the latent state with its hardest child, $\delta_v$ for Pythia (left), Herwig (center) and their ratio (right), for the quark class in the quark-gluon task (full Lund plane). Central row: Lund density for Pythia, Herwig and their ratio. Bottom row: standard error on the mean of $\delta_v$ for Pythia, Herwig and their ratio.}
    \label{fig:q_full_density}
\end{figure*}

\begin{figure*}[htpb]
    \centering
    \includegraphics[width=0.70\textwidth]{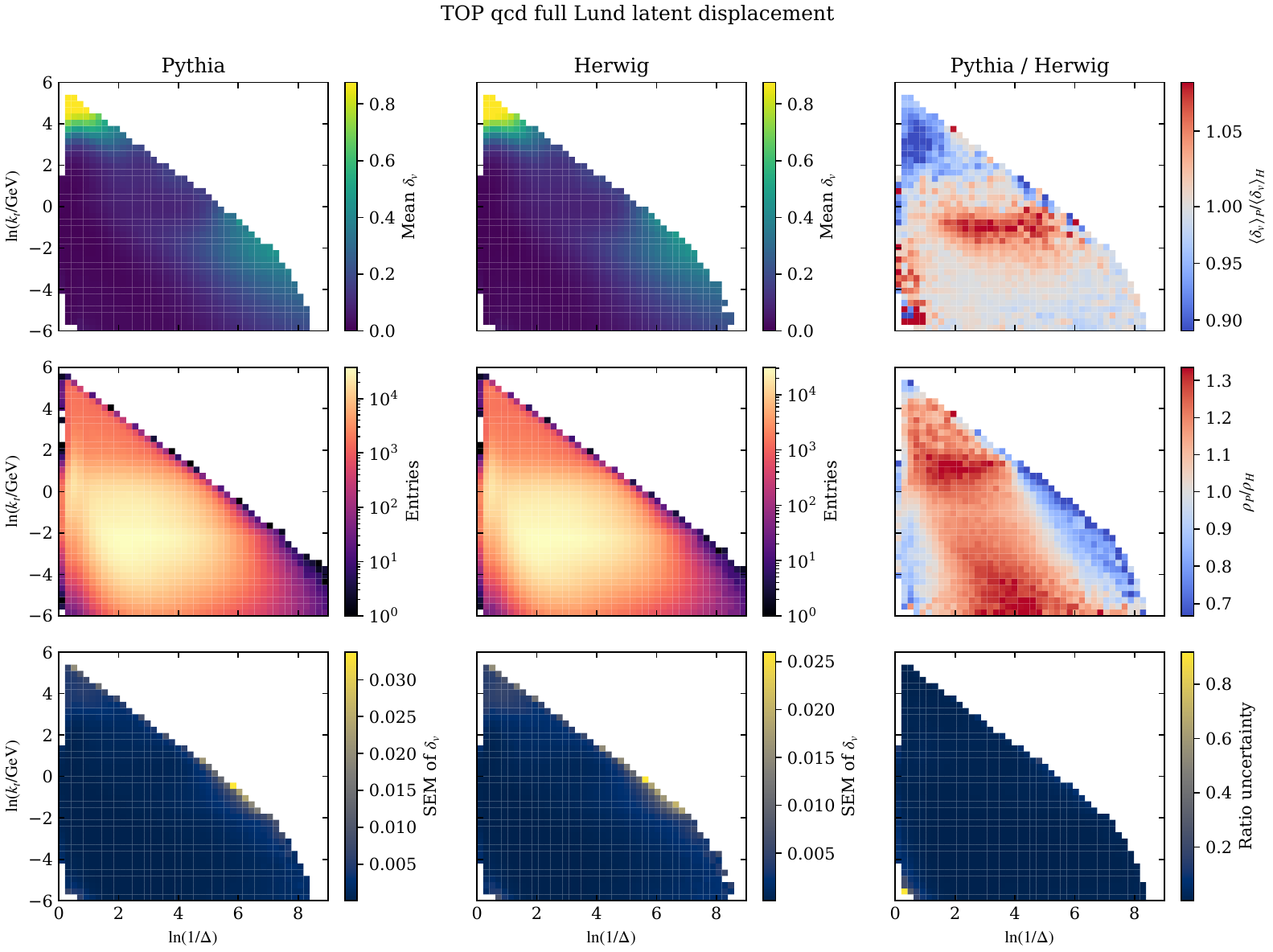}
    \caption{Top row: mean cosine distance of the latent state with its hardest child, $\delta_v$ for Pythia (left), Herwig (center) and their ratio (right), for the QCD class in the top tagging task (full Lund plane). Central row: Lund density for Pythia, Herwig and their ratio. Bottom row: standard error on the mean of $\delta_v$ for Pythia, Herwig and their ratio.}
    \label{fig:qcd_full_density}
\end{figure*}

\begin{figure*}[htpb]
    \centering
    \includegraphics[width=0.70\textwidth]{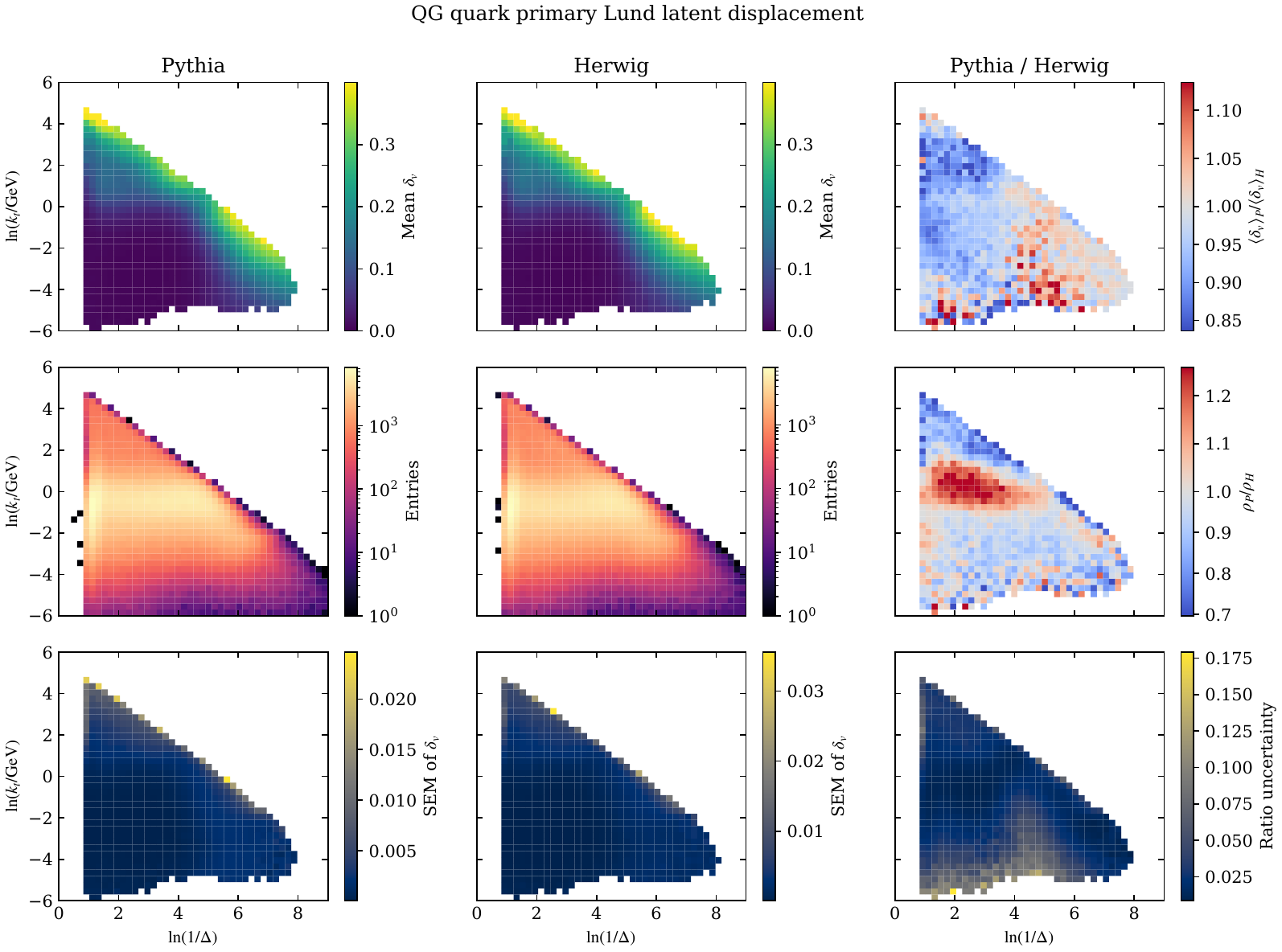}
    \caption{Top row: mean cosine distance of the latent state with its hardest child, $\delta_v$ for Pythia (left), Herwig (center) and their ratio (right), for the quark class in the quark-gluon task (primary Lund plane). Central row: primary Lund density for Pythia, Herwig and their ratio. Bottom row: standard error on the mean of $\delta_v$ for Pythia, Herwig and their ratio.}
    \label{fig:q_primary_density}
\end{figure*}

\begin{figure*}[htpb]
    \centering
    \includegraphics[width=0.70\textwidth]{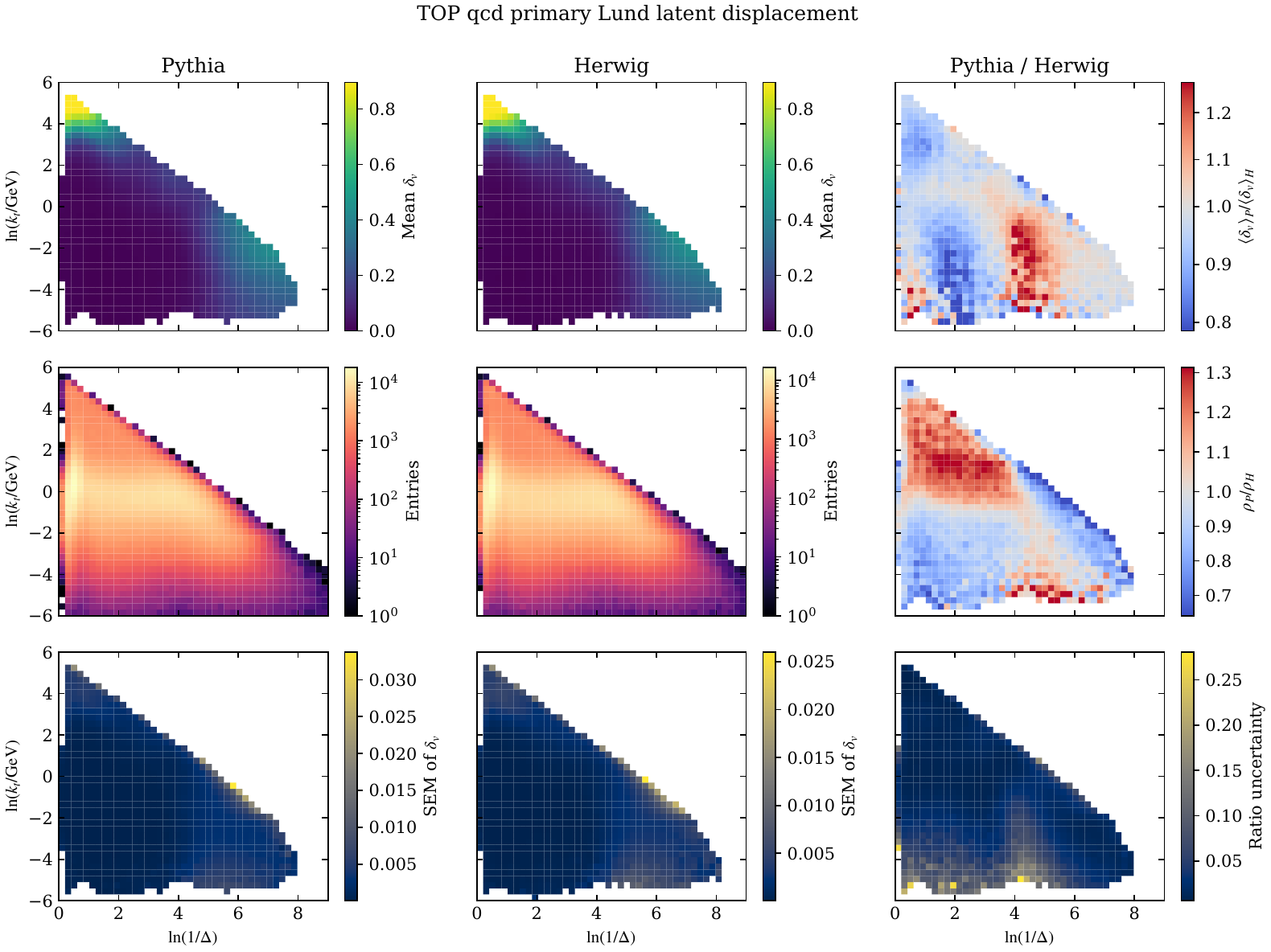}
    \caption{Top row: mean cosine distance of the latent state with its hardest child, $\delta_v$ for Pythia (left), Herwig (center) and their ratio (right), for the QCD class in the top tagging task (primary Lund plane). Central row: primary Lund density for Pythia, Herwig and their ratio. Bottom row: standard error on the mean of $\delta_v$ for Pythia, Herwig and their ratio.}
    \label{fig:qcd_primary_density}
\end{figure*}

\bibliographystyle{iopart-num} 

\bibliography{references.bib}        

\end{document}